\documentclass[reprint,amsmath,amssymb,aps]{revtex4-1}
\usepackage{graphicx}% Include figure files
\usepackage{dcolumn}% Align table columns on decimal point
\usepackage{bm}% bold math
\usepackage{amsmath}
\usepackage{dsfont}
\usepackage{subcaption}
\usepackage{braket}
\usepackage{epstopdf}
 \usepackage{xcolor}	
 \usepackage{changepage}
\usepackage{subcaption}
\usepackage{graphicx}
\usepackage{hyperref}
\usepackage{txfonts}
\usepackage{natbib}
\hypersetup{
hypertexnames = false,
colorlinks = true,
urlcolor = blue,
linkcolor = blue,
citecolor = blue
}
\def\ps@pprintTitle{%
 \let\@oddhead\@empty
 \let\@evenhead\@empty
 \def\@oddfoot{}%
 \let\@evenfoot\@oddfoot}
\makeatother
\usepackage{subcaption}
\usepackage{fancyhdr}

\begin{document}

\title{Generation of Structured Light and Controlled-NOT Gate in Microwave Regime}% Force line breaks with \\
\author{Parkhi Bhardwaj}
\email{parkhi.21phz0013@iitrpr.ac.in}
\author{Shubhrangshu Dasgupta}
\affiliation{
Department of Physics, Indian Institute of Technology Ropar, Rupnagar, Punjab 140001, India
}
\date{\today}
\begin{abstract}
%% Text of abstract
%difference frequency generation and the creation of vortex beams and Hollow Gaussian Beams in the microwave domain.
We propose how to generate beams with non-zero orbital angular momentum in the microwave domain using atomic vapor medium and coherent control techniques. 
Our approach utilizes a difference frequency generation process in a centrosymmetric medium in the presence of a dc electric field for frequency conversion and parametric amplification. By employing phase matching conditions and orbital angular momentum conservation, we constructed a Controlled NOT gate using a three-level atomic configuration. By generating Laguerre-Gaussian fields in the microwave domain, we open up novel possibilities for advanced information processing in wireless communication and potential applications in quantum technologies.
%A significant aspect of our study was the ability to manipulate the beam's characteristics by controlling the intensity of the control field.
%The structuring of light holds immense significance in optics due to its wide range of applications, including optical tweezers and quantum information systems. 
%By controlling the properties of these specialized beams, we can unlock novel possibilities for precision manipulation and advanced information processing in wireless communication and quantum technologies.
\end{abstract}
%%Graphical abstract
%\begin{graphicalabstract}
%\includegraphics{grabs}
%\end{graphicalabstract}

%%Research highlights
%\begin{highlights}
%\item Research highlight 1
%\item Research highlight 2
%\end{highlights}

%\tableofcontents

%% \linenumbers

%% main text

%\pacs{}
\maketitle
\section{Introduction}
Electromagnetic (EM) waves possess energy and both linear and angular momenta. The spin angular momentum (SAM) is associated with the circularly polarized
light while the orbital angular momentum (OAM) is related to the spatial phase
structure of light. Allen {\it et al.} discovered that light beams with an azimuthal phase
dependence of $\exp(il\phi)$ carry an OAM of $l\hbar$ per photon, where $\phi$ is the azimuthal coordinate in the
cross-section of the beam, and $l$ can take any positive or negative integer value \cite{r1,ALLEN1999291}.
The wavefront of the beam becomes helical in shape, with $l$ number of twists in one wavelength, while the Gaussian intensity profile displays a vortex (or null) at the center in the transverse direction of propagation. Such light beams can be generated by various methods, e.g., cylindrical lens pairs \cite{r3}, computer-generated holograms \cite{r4}, spatial
light modulators \cite{r5}, liquid crystal spatial light modulators \cite{r6}, etc. A detailed review of these techniques can be found in \cite{r7}.

In the last few years, researchers have been primarily interested in the interaction
between atoms and light which have OAM.
The OAM of a field can be transferred to another field via light-matter interaction using certain protocols that depend upon the energy level configuration of the atoms \cite{r,ra,rb,rc,r8,r9,r10}. %For example,  the OAM transfer from a pump field to a probe field was suggested in a two-level system, thanks to a four-wave mixing (FWM) process \cite{r8}. A similar transfer was demonstrated also in both four-level tripod-type and double tripod-type atomic systems using two vortex beams \cite{r9}.
%Recently, the medium the transfer of the optical vortex in a four-level double-lambda scheme, through an FWM process, has been reported only by one vortex control field \cite{r10}. 
It was shown in a V-type atomic configuration that by introducing coherence between the excited states (i.e., in an electric dipole-forbidden transition)using a microwave field, it is possible to transfer the OAM from a field with Laguerre-Gaussian (LG) intensity profile to a plane wave \cite{r11}. A similar effect can be obtained in a four-level ladder-type configuration \cite{r12}.
The coherence in the excited states in a V-type configuration can also be induced via vacuum-induced coherence \cite{r13}, instead of a microwave field, and this can, as well, lead to the transfer of
the OAM from an LG beam to a weak planer field \cite{r11}. %Note that the strength of VICis nonzero for the systems which have nonorthogonal atomic-induceddipole moment vectors. 
The authors in \cite{r14} have extended their result of a $\Lambda$-type configuration to a generalized tripod system to show that an initial vortex state can be transformed into many, by preparing the atom in a superposition of the ground states. In a symmetry-broken ladder-type three-level quantum system, OAM transfer can be more efficiently done, via three-wave mixing,
in Autler-Townes splitting regime rather than in the electromagnetically induced transparency regime \cite{r15}. 

%For efficient transfer of OAM from one field to another field, it is required to reduce the absorption losses of light in an atomic medium. Vortex-controlled absorption in multilevel systems has attracted much attention in recent times. When a control vortex beam is applied, the EIT will disappear because of the absorption losses at the vortex core \cite{r16,r17}.The EIT can reappear in the presence of either an extra nonvortex control laser beam or two control fields with opposite helicity \cite{r18}. The probe absorption spectrum can be manipulated by adjusting the OAM of the optical vortex beams and the frequency detunings in a four-level cascaded atomic system as well \cite{r19}. 

Until now, most of the research has focused on generating LG beams in the optical domain. However, due to the rapid advancement of wireless communication technology and an increasing demand for broadband data transfer, data centers, and cloud-based services \cite{r20}, there is a pressing need to enhance spectral efficiency and system capacity, in the microwave regime and beyond. 
To address this challenge, fields in the microwave regime with OAM pose as a promising candidate for information transfer. The OAM offers additional degrees of freedom to simultaneously carry information at the same frequency within a single communication channel \cite{r21}, that too without increasing the frequency bandwidth \cite{r22}. Consequently, the utilization of OAM in wireless communications has become an area of intense research, especially in the microwave regime \cite{r23}, necessitating developing techniques for generating waves with microwave frequencies along with the OAM.
%Visible light, which enables us to perceive our surroundings, belongs to the vast electromagnetic spectrum comprising various other types of radiation. All electromagnetic radiations consist of transverse electric and magnetic waves that propagate at the speed of light, specifically in free space. These waves possess different frequencies and wavelengths. Among them, microwaves are a form of electromagnetic radiation characterized by wavelengths ranging from $10^{-3}$ to $10^{-1}$ meters and frequencies spanning from 3x$10^9$ to 3x$10^{11}$ Hz \cite{D H Staelin}. As a result, microwaves possess higher energy, making them suitable for carrying signals over long distances with minimal signal degradation, or attenuation. Cost-effectiveness is one of the most prominent peculiarities that makes microwaves a preferred option for communication in the EM spectrum. LG beams are highly advantageous due to their numerous degrees of freedom. Consequently, LG beams in the Microwave regime become even more useful in communication as they possess higher energy and enable lossless transmission over long distances. \\

Microwave radiation cannot be generated via natural atomic transitions alone and so, we must manipulate the atomic energy levels. In this context, we use the Stark effect \cite{r29} and second-order nonlinear processes \cite{r31,r32,rd} in atomic vapors to generate microwave radiation. Even-order nonlinear processes are typically prohibited in centrosymmetric media under the dipole approximation. However, they can become significant if electric quadrupole and magnetic dipole matrix elements are included \cite{r27}, though these effects are generally weak and difficult to measure. Second-order nonlinearity can be observed in atomic vapor subjected to an electric or magnetic field, which breaks the symmetry by mixing states of even and odd parity \cite{r29, r26}.
Here, we propose a novel method for generating an LG beam in the microwave regime. Our model relies on three-wave mixing, which becomes feasible in an atomic quantum system excited by collinear lasers in the presence of a DC electric field. We carefully calibrate the electric field intensity to ensure an energy splitting that surpasses the fine structure splitting, setting it at approximately 10$^{3}-10^{4}$ V/cm \cite {r29}. This setup enables a cyclic energy level configuration to transfer the OAM from the input field to the output field. 

In this paper, we demonstrate, for the first time, the generation of a field in the microwave regime using optical fields and discuss the efficiency of this process.
Subsequently, we introduce phase singularity on the generated field through the coupling and control fields, individually selecting them as LG beams. We then explore how the intensity of the generated field can be reshaped by manipulating the strength of the control field. Finally, we successfully generate the hollow Gaussian beam (HGB) by incorporating the Gaussian profile of the control field.
Then we discuss the construction of a controlled NOT (CNOT) gate in the OAM basis using three-level atomic energy level configuration. 

The structure of the paper is as follows.
In Sec. II, we show how a V-type configuration can lead to microwave generation. The study of a $\Lambda$ configuration in the same context is described in Sec. III. How a CNOT gate can be created in these configurations is discussed in Sec. IV. In Sec. V, we propose ways of creating dipole-allowed transitions to facilitate microwave generation. We conclude the paper in Sec. VI. 

%In this article firstly, we use a V-type system to generate a microwave field, and using to LG field in an optical regime, we transfer their OAM coherently to the generated Microwave field. Secondly, we have done the same using a $\Lambda$-type system.

\section{Microwave generation with V-type configuration}
\begin{figure}[ht!]
\centering
\begin{subfigure}{.45\textwidth}
    \includegraphics[width=\textwidth, height=5.5 cm]{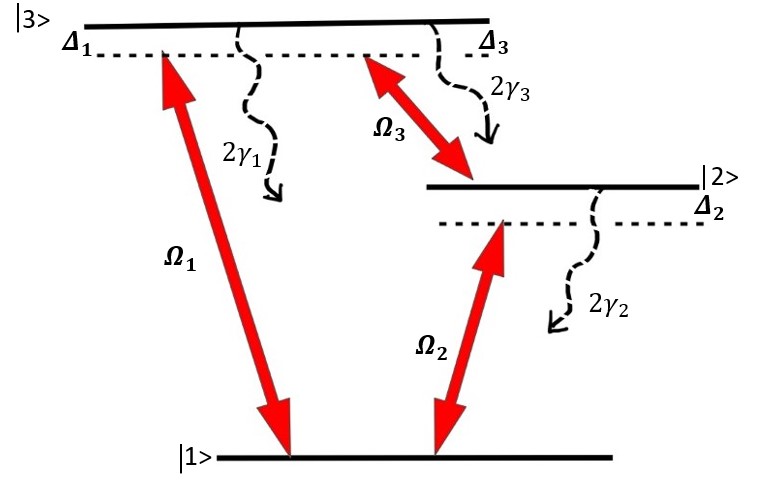}
    \caption{}
    \label{fig1}
\end{subfigure}
\hfill
\begin{subfigure}{.45\textwidth}
    \includegraphics[width=\textwidth, height=4 cm]{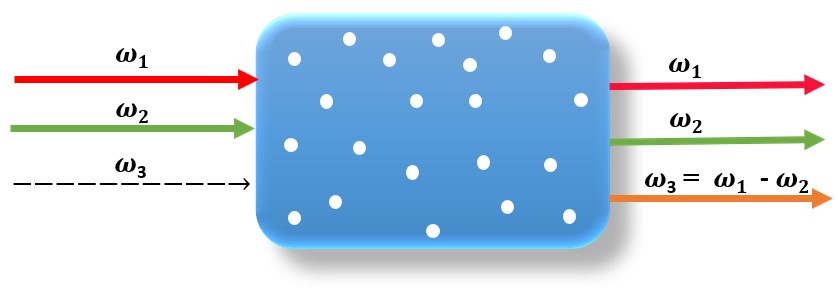}
    \caption{}
    \label{4a}
\end{subfigure}
\caption{(a) Relevant energy level configuration of the  V-type three-level atomic system. (b) A schematic to depict the three wave mixing in atomic vapor}
\end{figure}

We consider a V-type energy level configuration of the atomic system as illustrated in Fig. \ref{fig1}. The symmetry of the system can be disrupted by applying a dc electric field in the $z$-direction, resulting in the mixing of energy levels of even and odd parity. The frequency gap between the energy levels can be adjusted by the applied dc electric field. 

To generate a microwave field, we apply a dc electric field on the atom that splits the energy levels, mixes the energy levels of even and odd parity, and creates a gap in the order of the microwave frequency domain between the energy levels of the same principle quantum number, according to the strength of the electric field. 
% Now we present a general energy level configuration that allows the cyclic transition. 
The transitions to the excited states $|2\rangle$ and $|3\rangle$ from the ground state $|1\rangle$ are induced by two electromagnetic fields with amplitudes $\vec{E}_1$ and $\vec{E}_2$ and frequencies $\omega_1$ and $\omega_2$, respectively.
%The transition $|2\rangle\leftrightarrow |3\rangle$ is magnetic-dipole allowed and is driven by a magnetic field of frequency $\omega_3$. 
The transitions $\ket{1}\leftrightarrow\ket{3}$, $\ket{3}\leftrightarrow \ket{2}$, and $\ket{1}\leftrightarrow \ket{2}$ are electric dipole-allowed. 
 with respective electric dipole moments $\vec{\mu}_{31}$, $\vec{\mu}_{32}$  and $\vec{\mu}_{21}$. The Rabi frequencies of the applied fields are defined as $\Omega_1=\frac{\vec{\mu}_{31}.\vec{E_1}}{\hbar}$ and $\Omega_2=\frac{\vec{\mu}_{21}.\vec{E_2}}{\hbar}$, %and $\Omega_3=\frac{\overrightarrow{\mu}_{32}.\overrightarrow{B}}{\hbar}$. 
 %and $\overrightarrow{E}_1$, and $\overrightarrow{E}_2$ are the amplitude of the  electric field, 
%and  $\overrightarrow{B}$ is the amplitude of magnetic dipole moment. 
This energy level configuration can be observed in hydrogen-like alkali atoms in the presence of a dc electric field. %Additionally, similar configurations can be demonstrated in sodium, cesium, and rubidium atoms in the presence of a dc electric field. 
This quantum system exhibits second-order nonlinear wave-mixing because of the lack of inversion symmetry, leading to the generation of a new frequency $\omega_3 = \omega_1 - \omega_2$, via difference frequency generation. This requires the following phase matching condition: $\vec{k_3} = \vec{k_1} - \vec{k_2}$. The three-wave phase matching condition allows for coupling between the two excited states, with the transition occurring in the microwave frequency regime.

We assume that one of these electromagnetic fields in the visible regime possesses the LG intensity profile, along with a non-zero OAM. The respective Rabi frequencies for both the fields are given by (in cylindrical coordinates)

\begin{equation}
\Omega_j(r,\phi)= \Omega_{0j}\frac{1}{\sqrt{|l_j|!}}\left(\frac{\sqrt{2}r}{w}\right)^{|l_j|} e^{-r^2/w^2}e^{i l_j\phi}\;,\;\; j\in 1,2\;,
\end{equation} 
where $\Omega_{0j}$, $l_j$ and $\omega$ define the constant Rabi frequency, the topological charge of the $j$th field, and beam waist respectively.

The interaction between the atom and the EM fields is described by the following Hamiltonian, under electric-dipole and rotating wave approximations (RWA): 
\begin{equation}
\begin{split}
 H=\hbar\omega_{21}\ket2\bra{2}+\hbar\omega_{31}\ket3\bra3- 
 (\hbar\Omega_1 e^{-i\omega_1t}\ket1\bra3 \\ +\hbar\Omega_2 e^{-i\omega_2t}\ket1\bra2+\hbar\Omega_3 e^{-i\omega_3t}\ket2\bra3 + h.c)\;,
 \end{split}
\end{equation}
where $\Omega_3$ is the Rabi frequency of the generated signal field. 
In order to find the dynamical behavior of the atomic population and coherences of the V-type system, we use the Liouville equation:
\begin{equation}
    \dot{\rho} =-\frac{i}{\hbar}[H, \rho] + L\rho,
\end{equation}
where the second term describes the spontaneous decay processes of the atomic system. Then using the Liouville equation, the density matrix equations under the RWA  are given by
\begin{equation}
\begin{split} 
    \dot{\Tilde{\rho}}_{22} = i\Omega^*_2\Tilde{\rho}_{12} -i\Omega_2\Tilde{\rho}_{21} + i\Omega_3\Tilde{\rho}_{32} -i\Omega^*_3\Tilde{\rho}_{23}  \\ -2\gamma_{12}\Tilde{\rho}_{22} + 2\gamma_{32}\Tilde{\rho}_{33} \\
\dot{\Tilde{\rho}}_{33} = i\Omega^*_1\Tilde{\rho}_{13} - i\Omega_1\Tilde{\rho}_{31} - i\Omega_3\Tilde{\rho}_{32} +i\Omega^*_3\Tilde{\rho}_{23}  \\ -2(\gamma_{13}+ \gamma_{32})\Tilde{\rho}_{33}\\ 
\dot{\Tilde{\rho}}_{12} = -i\Omega_2(\Tilde{\rho}_{22} - \Tilde{\rho}_{11}) + i\Omega_1\Tilde{\rho}_{32} - i\Omega^*_3\Tilde{\rho}_{13} \\+  
 (i\Delta_2 - \gamma_{12})\Tilde{\rho}_{12}\\
\dot{\Tilde{\rho}}_{13} = -i\Omega_1(\Tilde{\rho}_{33} - \Tilde{\rho}_{11}) +i\Omega_2\Tilde{\rho}_{23} - i\Omega_3\Tilde{\rho}_{12}  \\ +(i\Delta_1 - (\gamma_{13}+\gamma_{32}))\Tilde{\rho}_{13}\\
\dot{\Tilde{\rho}}_{23} = -i\Omega_1\Tilde{\rho}_{21} + i\Omega^*_2\Tilde{\rho}_{13} +i\Omega_3(\Tilde{\rho}_{22}-\Tilde{\rho}_{33})  \\+(i\Delta_3 - (\gamma_{13}+\gamma_{12}+
\gamma_{32}))\Tilde{\rho}_{23}.
\end{split}
\label{master}
\end{equation}
where $\Delta_1$ = $\omega_1$ - $\omega_{13}$, $\Delta_2$ = $\omega_2$ - $\omega_{12}$, and $\Delta_3$ = $\omega_3$ - $\omega_{32}$  are the detunings of fields. We assume that $E_1$ is the control field and $E_2$ is the coupling field. The above density matrix obeys the conditions $\sum_i\rho_{ii}$ =1
 and $\Tilde{\rho}_{ij}$ = $\Tilde{\rho}_{ji}^*$.
 To obtain the above equations we have used the following transformations: 
\begin{equation}
    \rho_{13} = \Tilde{\rho}_{13}e^{-i\omega_1t},\rho_{12} = \Tilde{\rho}_{12}e^{-i\omega_2t}, \rho_{23} = \Tilde{\rho}_{23}e^{-i\omega_3t}.
\end{equation}
In Eq. (\ref{master}), $\gamma_{13}$, $\gamma_{12}$, and $\gamma_{32}$ are the spontaneous decay rates of the states $|1\rangle$, $|2\rangle$ and $|3\rangle$ respectively.

In the weak-field limit, we can expand the density matrix elements up to first order in $\Omega_2$: $\Tilde{\rho}_{ij} =\Tilde{\rho}^{(0)}_{ij} + \Omega_2\Tilde{\rho}^{(1)}_{ij}+ ...$,
where $\Tilde{\rho}^{(k)}_{ij}$ are the $k$th order approximations. %and obtain the following analytical expression for the first order of $\Tilde{\rho}_{32}$ in the steady state.
%To calculate the coherence $\Tilde{\rho}_{32}$, we solve equation(4) for the steady state.
We assume that detunings $\Delta_1$, $\Delta_2$, and $\Delta_3$ are all zero, and the decay rates  $\gamma_{12}$ and $\gamma_{13}$ are equal to $\gamma$ , while $\gamma_{32}$ is equal to 2$\gamma$. This allows us to obtain the following expressions for the first order of $\Tilde{\rho}_{32}$ and $\Tilde{\rho}_{21}$ in the steady state:
\begin{equation}
 \Tilde{\rho}_{32}^{(1)} = \frac{-6\Omega_1\Omega_2^* }{18\gamma^2+|\Omega_1|^2}\hspace{.1cm},\hspace{.5cm}
 \Tilde{\rho}_{21}^{(1)} = \frac{-18i\gamma\Omega_2^*+6\Omega_1\Omega_3}{18\gamma^2+|\Omega_1|^2}\;. \label{den1}
 \end{equation}
 Here the real part of coherence  $\tilde{\rho}^{(1)}_{32}$ carries information about the beam profile, which is then transferred to the microwave field through interaction between light and matter. On the other hand, the imaginary part of the coherence $\tilde{\rho}_{32}$ corresponds to the amplification or absorption of the generated microwave field. If the imaginary part of coherence is negative 
 (positive), it corresponds to the gain or amplification (absorption). 
 
 We next study the beam propagation of the microwave field, generated through the atomic coherence $\tilde{\rho}_{32}$. For a time-independent microwave field propagating in the $z$ direction, the Maxwell's equation can be written in the slowly varying envelope approximation as
\begin{equation}
    \frac{\partial\Omega_3}{\partial z} = i\frac{\alpha_{32}\gamma_{32} }{2L}\hspace{.2cm} \Tilde{\rho}_{32}^{(1)}, \hspace{.5cm}
  \frac{\partial\Omega_2}{\partial z} = i\frac{\alpha_{21}\gamma_{21}}{2L}\Tilde{\rho}_{21}^{(1)} \;, 
  \label{max1}
\end{equation}
where  $\alpha_{32} =\alpha_{21} = \alpha$ is the optical depth and $L$ is the length of the medium. Substituting the Eq. (\ref{den1}) into Eq. (\ref{max1}) with initial condition $\Omega_3(z=0) = 0$ and $\Omega_2(z=0) = \Omega_2(0)$. We get the following expression for the Rabi frequency of the microwave field:
\begin{equation}
    \begin{split}
        \Omega_3 (z) = \frac{-4i\Omega_2^*(0)\Omega_1}{\sqrt{(9\gamma+8|\Omega_1|^2)}}\exp\left(\frac{9\alpha\gamma^2 z}{2L(18\gamma^2+|\Omega_1|^2)}\right) \\ 
    \sinh\left(\frac{3\alpha\gamma z\sqrt{9\gamma^2+8|\Omega_1|^2}}{2(18\gamma^2+|\Omega_1|^2)L}\right)\;.
    \label{o3}
     \end{split}
\end{equation}

%We consider two scenarios. Firstly, when the first input field is weak, it becomes evident from Fig(2) that the second input field is also weak to facilitate the efficient generation of the third field. Conversely, if the first field is strong, the intensity of the generated field will increase exponentially for the strong second input field. For these plots, we conclude that for efficient generation of microwave fields both input fields are weak or both are strong.
%\begin{figure}
     %\centering
     %\begin{subfigure}[b]{.7\textwidth}
        % \centering
        % \includegraphics[width=\textwidth]{v(o1=.5).jpg}
        % \caption{}
        % \label{fig:y equals x}
     %\end{subfigure}
    % \hfill
     %\begin{subfigure}[b]{.7\textwidth}
        % \centering
         %\includegraphics[width=\textwidth]{v(o1=5).jpg}
        % \caption{}
         %\label{fig:three sin x}
    %\end{subfigure}
        %\caption{Intensity of $\Omega_3$ against propagation distance for two cases in (a) $\Omega_1$ is 0.5$\gamma$ and in (b) $\Omega_1$ is 5$\gamma$. }
        %\label{fig:three graphs}
%\end{figure}
%\begin{figure}[h]

		%\centering
		%\includegraphics[scale=.8]{modeomega3vsZ.png}
		%\caption{Intensity of $\Omega_3$ against propagation distance for two cases in (a) $\Omega_1$ is 0.5$\gamma$ and in (b) $\Omega_1$ is 5$\gamma$. } 
%\end{figure}
%To know the strength of the input fields for which we get the generated field without much losses into the medium, we plot the $\Omega_3$ as a function of input fields
%\begin{figure}[h]
		%\centering
		%\includegraphics[scale=0.2]{v(o3 vs o1).jpg}
		%\caption{$\Omega_3$ as a function of $\Omega_1$ for $\Omega_2 = 0.5$\gamma$ and 5$\gamma$, (Z/L) = 1.}
	%\end{figure}

 \subsection{Efficiency of Three-wave mixing}
 To evaluate the efficiency of the above process of a third field through three-wave mixing, we define a dimensionless quantity $\eta = (|\Omega_3(z)|/|\Omega_2(0)|)^2$ \cite{r25}, given by
 \begin{equation}
    \begin{split}
        \eta = \frac{16\Omega_1}{(9\gamma+8|\Omega_1|^2)}\exp\left(\frac{9\alpha\gamma^2 z}{L(18\gamma^2+|\Omega_1|^2)}\right) \\ 
    \sinh^2\left(\frac{3\alpha\gamma z\sqrt{9\gamma^2+8|\Omega_1|^2}}{2(18\gamma^2+|\Omega_1|^2)L}\right)\;.
     \end{split}
\end{equation}

\begin{figure*}
\centering
\begin{subfigure}{.49\linewidth}
    \includegraphics[width=8cm]{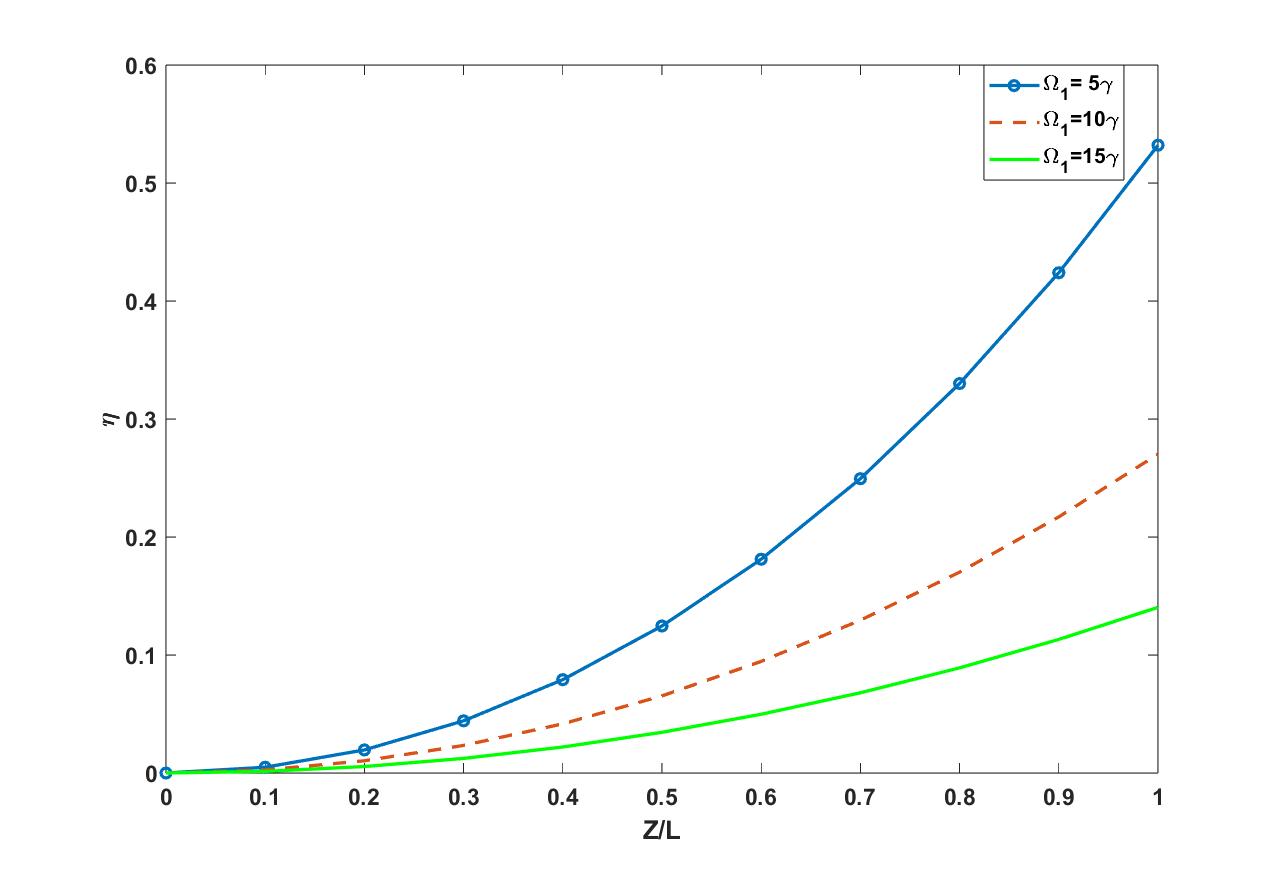}
    \caption{}
    \label{2a}
\end{subfigure}
\hfill
\begin{subfigure}{.49\linewidth}
    \includegraphics[width= 8cm]{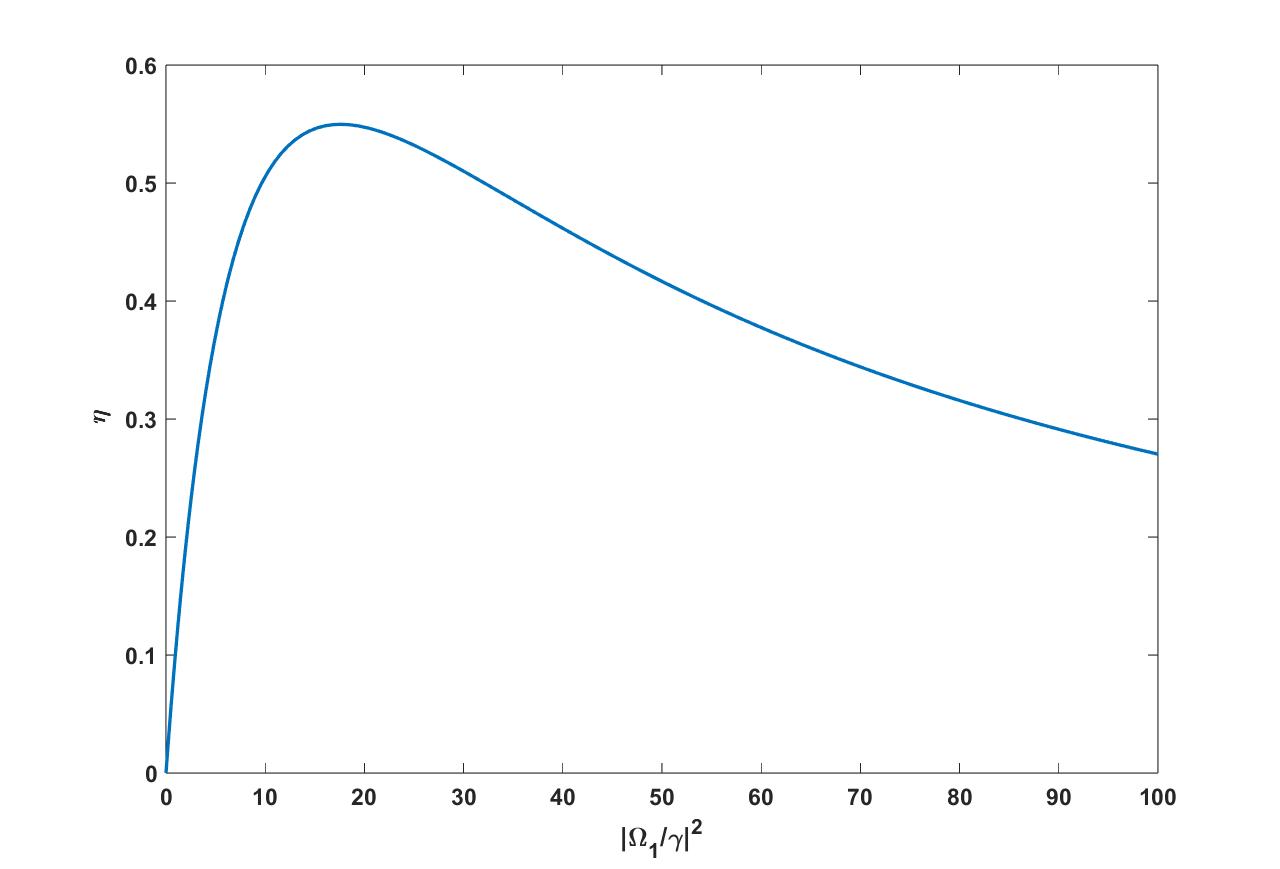}
    \caption{ }
    \label{2b}
\end{subfigure}
\hfill
\begin{subfigure}[t]{1\linewidth}
    \includegraphics[width=9cm]{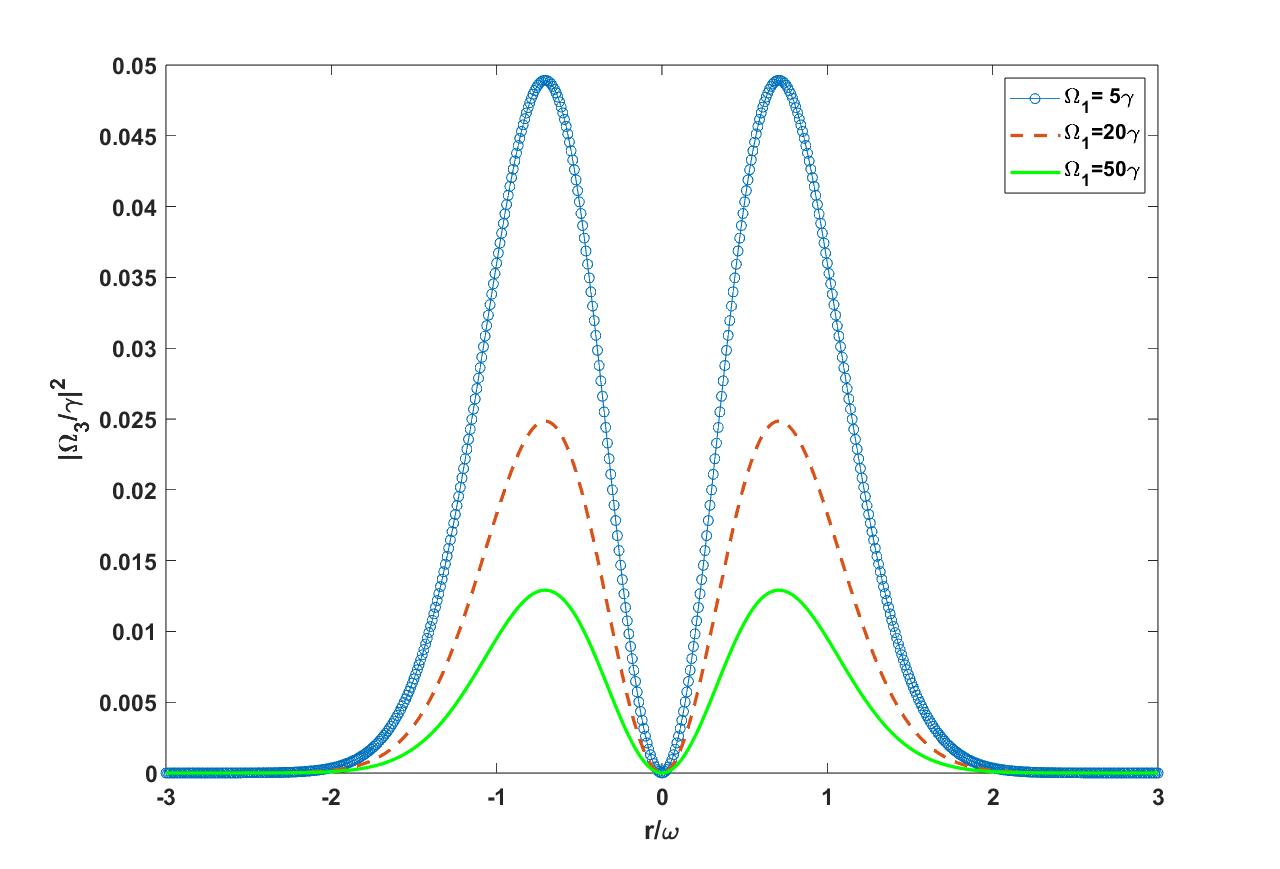}
    \caption{ }
    \label{2c}
\end{subfigure}
\caption{Variation of (a) $\eta$ as a function of $Z/L$ for $\Omega_1 = 5\gamma$ (blue (solid-o)), $\Omega_1= 10\gamma$ (red (dash)), and $\Omega_1 = 15\gamma$ (green (solid)), (b) $\eta$ as a function of $|\Omega_1/\gamma|^2$, and (c) $(|\Omega_3|/\gamma)^2$ with respect to $r/\omega$ for $\Omega_1 = 5\gamma$ (blue (solid-o)), $\Omega_1= 20\gamma$ (red (dash)), and $\Omega_1 = 50\gamma$ (green (solid)). We have chosen  $\Omega_{02}=0.5\gamma$, $l_2=1$, and $Z = L$.}
\end{figure*}

We plot in Fig. \ref{2a}, $\eta$ as a function of the parameter $Z/L$, where, $Z=z\alpha$ is the modified propagation distance, for various strengths of the field $\Omega_1$. 
As shown in this figure, the efficiency of three-wave mixing begins to increase with increasing $Z/L$. Since nonlinear processes are based on momentum and energy conservation \cite{r28}, the strong control field $\Omega_1$ leads to the amplification of the coupling field $\Omega_2$ and the generated signal field $\Omega_3$. Since the medium in consideration exhibits second-order nonlinearity, the interaction of $\Omega_1$ and $\Omega_2$ generates a signal field that gets amplified as it keeps propagating further into the medium.
After traveling a certain distance into the medium, a generated field emerges with its amplitude, dependent on the strength of the control field. This process involves frequency conversion with gain, referring to the parametric amplification of the generated field. The efficiency factor $\eta$ exhibits exponential growth with respect to the $Z$. Therefore, the gain of the generated field depends upon both the optical depth $\alpha$ and the strength of the control field $\Omega_1$. While it may be challenging to dynamically control the atomic number density of the medium (and hence, $\alpha$), we can monitor $\Omega_1$ to achieve the desired outcome.
We show in Fig. \ref{2b} the variation of the efficiency $\eta$ with $(|\Omega_1|/\gamma)^2$. It is clear from this figure that there is an optimal value of $\Omega_1$, for which the $\eta$ is maximum. 

\subsection{Transfer of OAM from the coupling field to the generated field}
Next we consider transferring the OAM of a coupling field to the generated field. From the Eq. \ref{o3}, we can see that the $\Omega_3$  is proportional to $\Omega^*_2(0)$. This means that a field of topological charge, \(l_2\) can be formed if the coupling field contains a charge \(-l_2\) at $z=0$. We show in the Table \ref{1}, the intensity and phase profile of $\Omega_3$ and the imaginary part of $\rho_{32}$ at a propagation distance $Z=L$ for $l_2 = 1, 2,$ and 3. As we increase the value of $l
_2$, the area of the null point of intensity also increases and the phase changes accordingly: for $l_2 = 1$ it changes from 0 to $2\pi$; for $l_2=2$, from 0 to 4$\pi$, and for $l_2=3$, from 0 to 6$\pi$. If $l_2$ is the topological charge of the generated field, one can see 2$l
_2$ numbers of petals in the third column of Table \ref{1}, in which the positive (negative) values of ${\rm imag}(\rho_{32}$ corresponds to the absorption (gain) of the generated field.  We also show in Fig.\ref{2c} the radial distribution of the intensity of the generated field with topological charge $l_2=1$ for $\Omega_1$ = 5$\gamma$, 10$\gamma$, and 15$\gamma$. We find that for larger $\Omega_1$,  the intensity of the generated field gets reduced. 

\begin{table*}[ht!]
     \begin{center}
     \begin{tabular}{ |p{.7cm} | p{15cm} | }
     \hline
      ( $l_2)$ & \hspace{3cm}Intensity \hspace{3cm} Phase \hspace{3.5cm}     Im($\rho_{32})$\\ \hline
      1 & \hspace{0cm} \includegraphics[width=.8\textwidth, height=45mm]{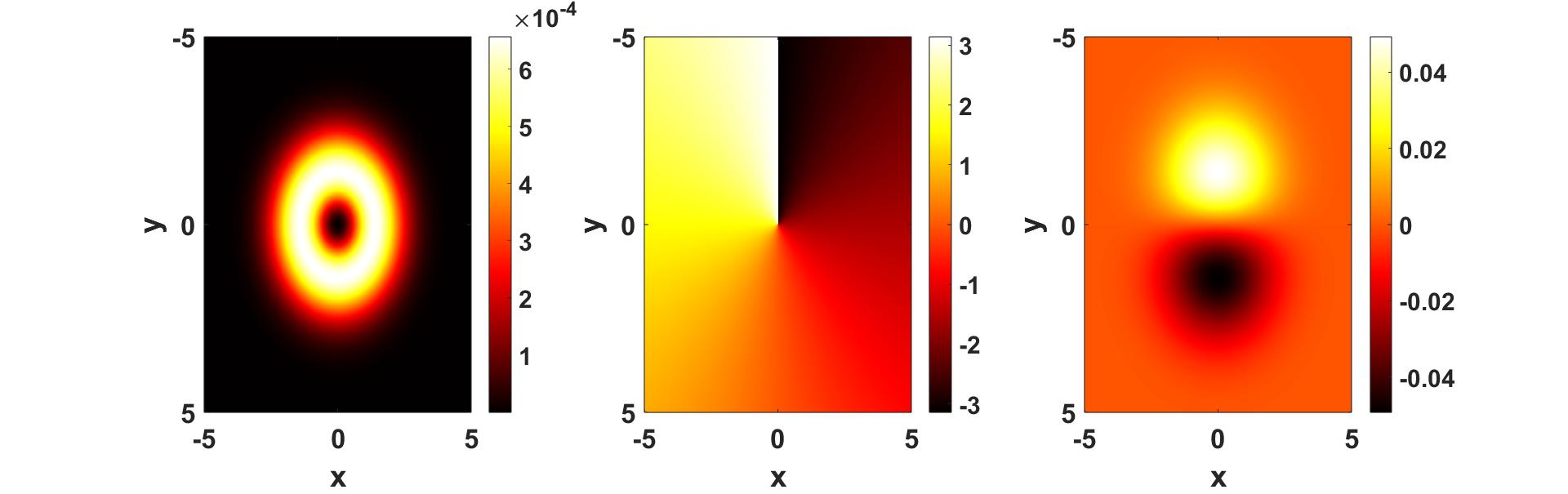}
      \\ \hline
      2 &  \includegraphics[width=0.8\textwidth, height=45mm]{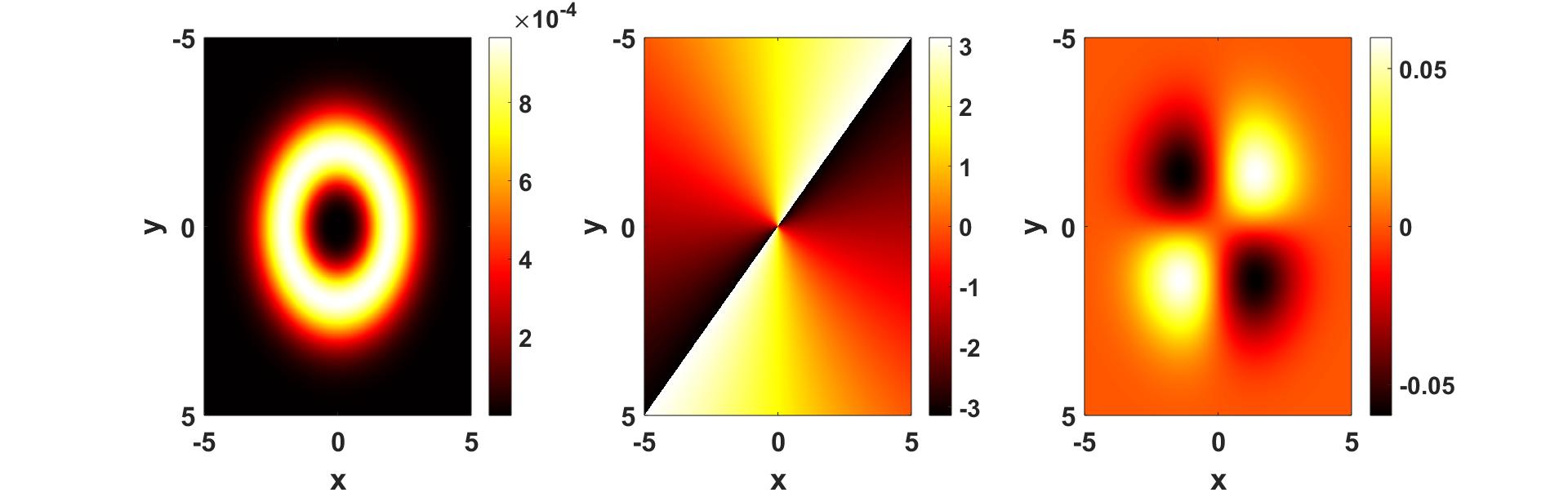}
      \\ \hline
      3 &\includegraphics[width=0.8\textwidth, height=45mm]{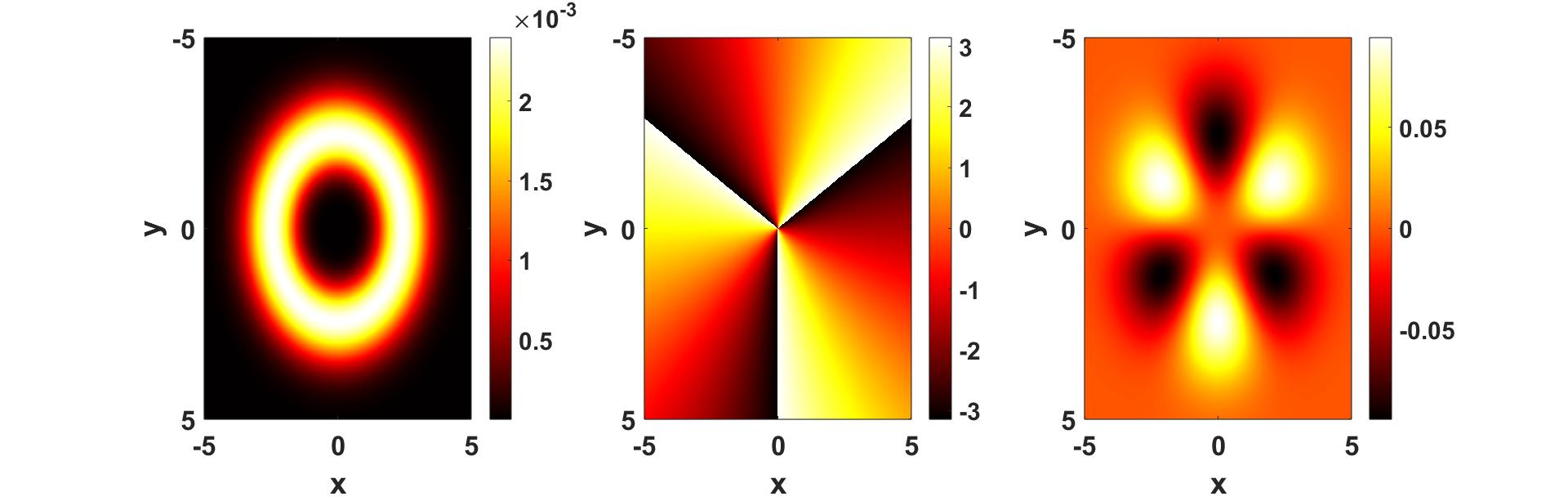}
      \\ \hline
    
      \end{tabular}
      
      \caption{Intensity (first column), phase profile (second column) and,  imag($\rho_{32}$ ) of the generated field of rabi frequency $\Omega_3$ for different modes of coupling LG field, i.e, $l_2$ = 1, 2 and, 3. The parameters are $\Omega_1=5\gamma$, $\Omega_{02}=0.5\gamma$, and $Z=L$.}
      \label{1}
      \end{center}
        
      \end{table*}

\subsection{Transfer of OAM from the Control Field to the Generated Field}

\begin{figure}
\begin{subfigure}{1\linewidth}
    \includegraphics[width=8cm]{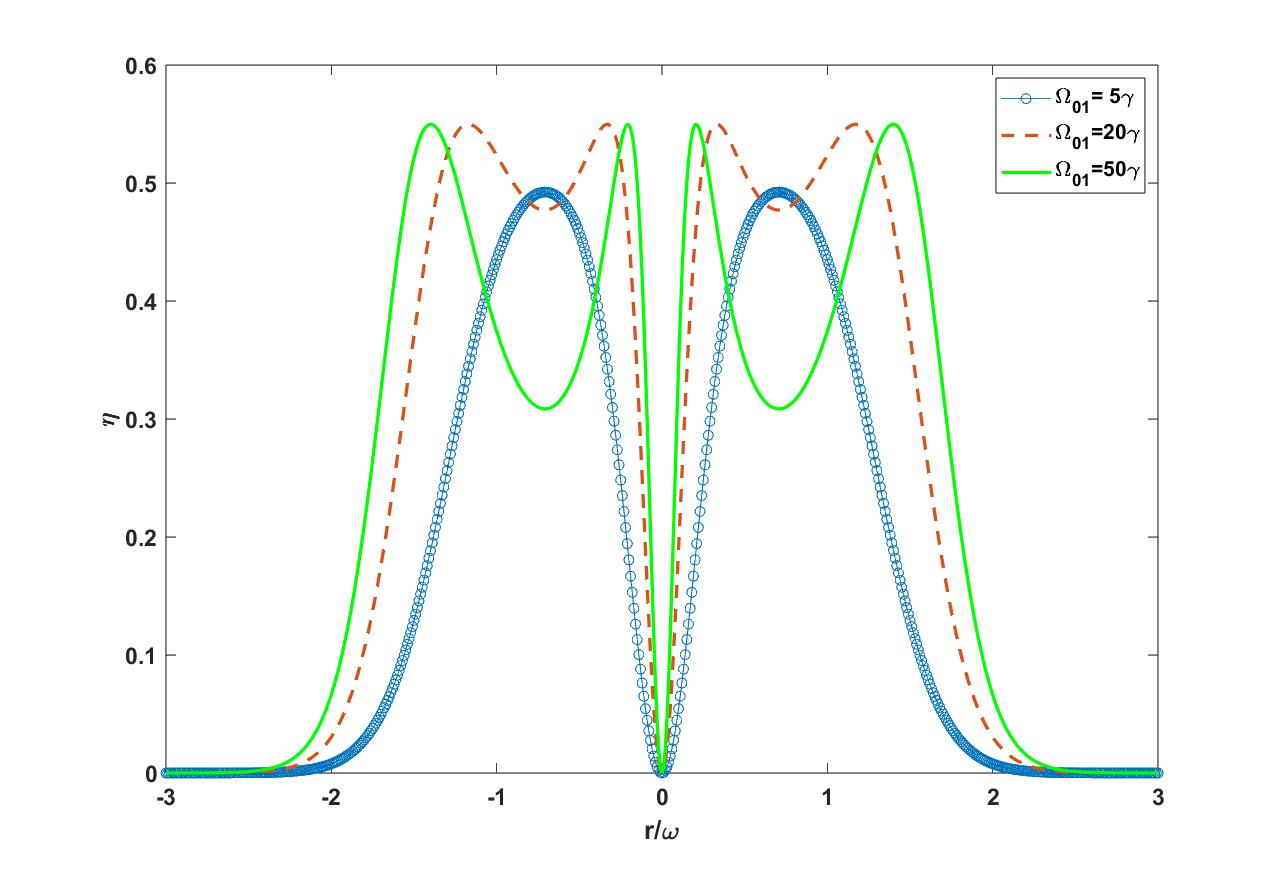}
    \caption{}
    \label{3a}
\end{subfigure}
\hfill
\begin{subfigure}{1\linewidth}
    \includegraphics[width=8cm]{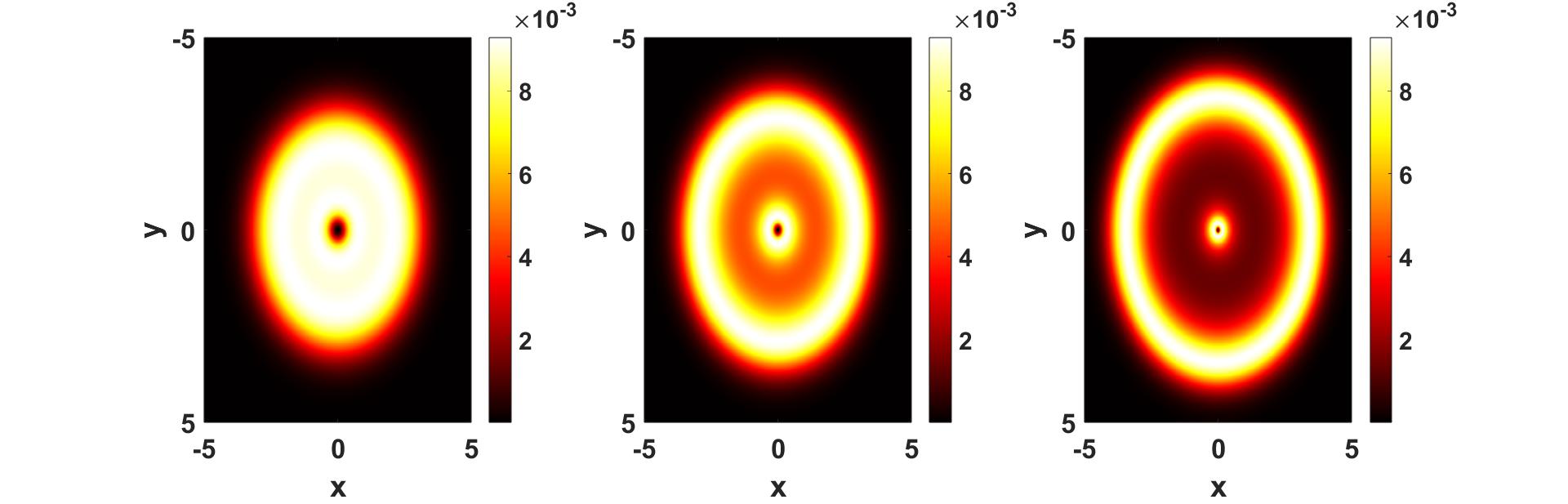}
    \caption{}
    \label{3b}
\end{subfigure}
\caption{(a) Variation of the $\eta$ as a function of $r/w$ when $\Omega_{01}=5\gamma$ (blue, circled dots), 20$\gamma$ (red, dash), and 50$\gamma$ (green, solid); (b) intensity profile of the generated vortex beam when $\Omega_{01}=5\gamma$, 20$\gamma$, and 50$\gamma$, respectively, and the other parameters $\Omega_2= 0.5\gamma$, and $\omega$=2 mm. We have chosen $l_3$=1 at $Z=L$.}
\end{figure}

So far, we treated the coupling field as an LG field. Since the generated field $\Omega_3$ is also proportional to $\Omega_1$ [see Eq. \ref{o3}], the phase singularity of the control field $\Omega_1$  can also be transferred to the generated field. Now, we are considering the control field to be the LG field. To understand the intensity pattern, we show in Fig. \ref{3a} the variation of $\eta$ with respect to $r/w$, for different values of $\Omega_{01}$. This plot illustrates that one ring of the generated LG field with  $l_3=1$ starts to split into two rings around the area of null intensity as we increase the strength of the control field. In Fig. \ref{3b}, we display the intensity profile of the generated field for different values of $\Omega_{01}$. We see appearances of two concentric rings of intensity (resembling two doughnuts),  for larger $\Omega_{01}$.  
%As we increase the value of $\Omega_{01}$, the ring begins to split into multiple rings. By appropriately selecting the strength of the control field, we can shape our beam according to our requirements.\\

\subsection{Generation of Hollow Gaussian Beam}
As we have seen in the preceding subsections, just by monitoring the strength of the control field, one can shape the generated light and split a single doughnut into two, which form an LG beam with azimuthal mode $l_3$. If the intensity profile of the control field is Gaussian ($l_1$=0), the lower values of $\Omega_{01}$ yield a Gaussian intensity pattern for generated field. However, as we progressively increase $\Omega_{01}$, the generated beam splits into ring, forming a hollow Gaussian beam. We show in Fig.\ref{4a} the normalized intensity of the generated field for different values of $\Omega_{01}$. As we increase $\Omega_{01}$, the intensity null at $r/\omega=0$ becomes prominent.

\begin{figure}[ht!]
\begin{subfigure}{1\linewidth}
    \includegraphics[width=8cm]{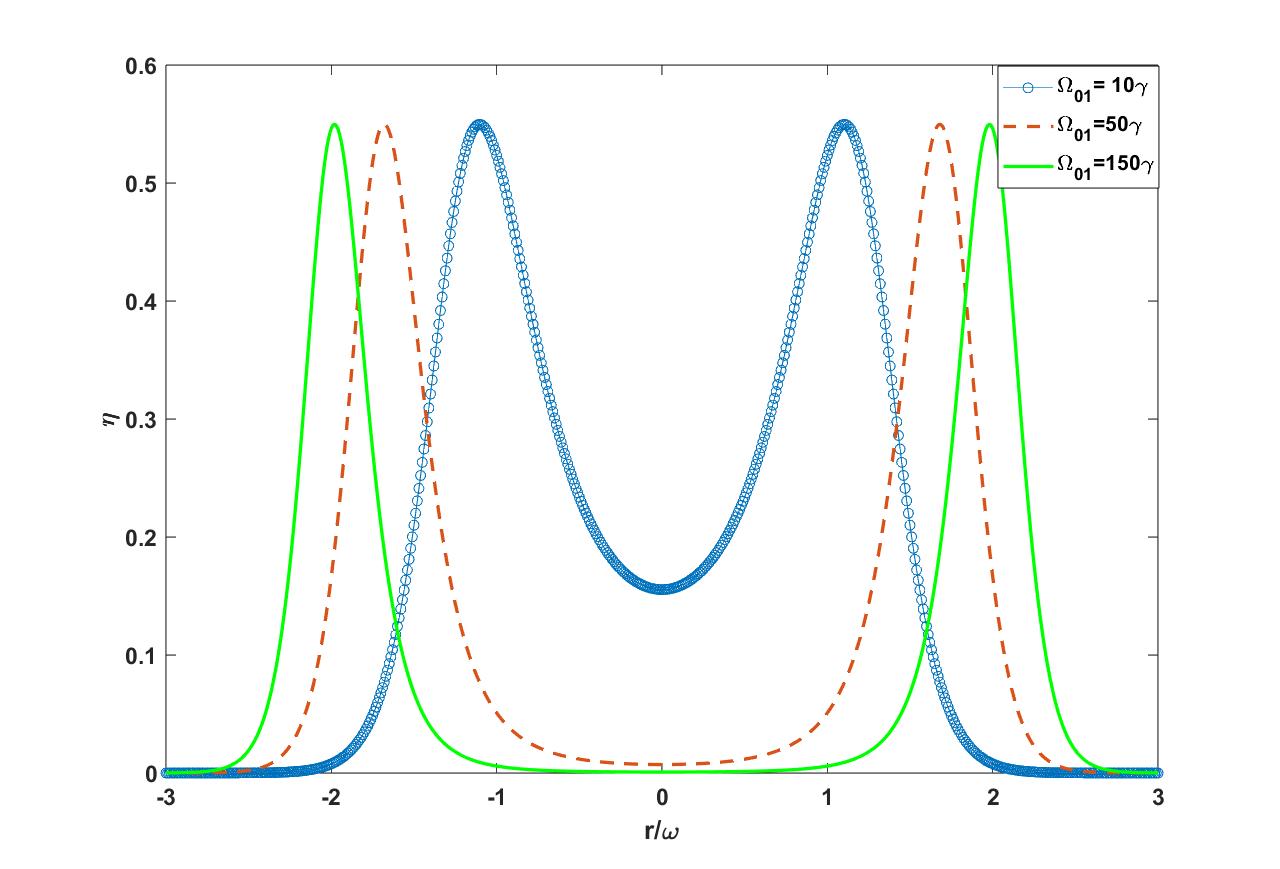}
    \caption{}
    \label{4a}
\end{subfigure}
\hfill
\begin{subfigure}{1\linewidth}
    \includegraphics[width=8cm ]{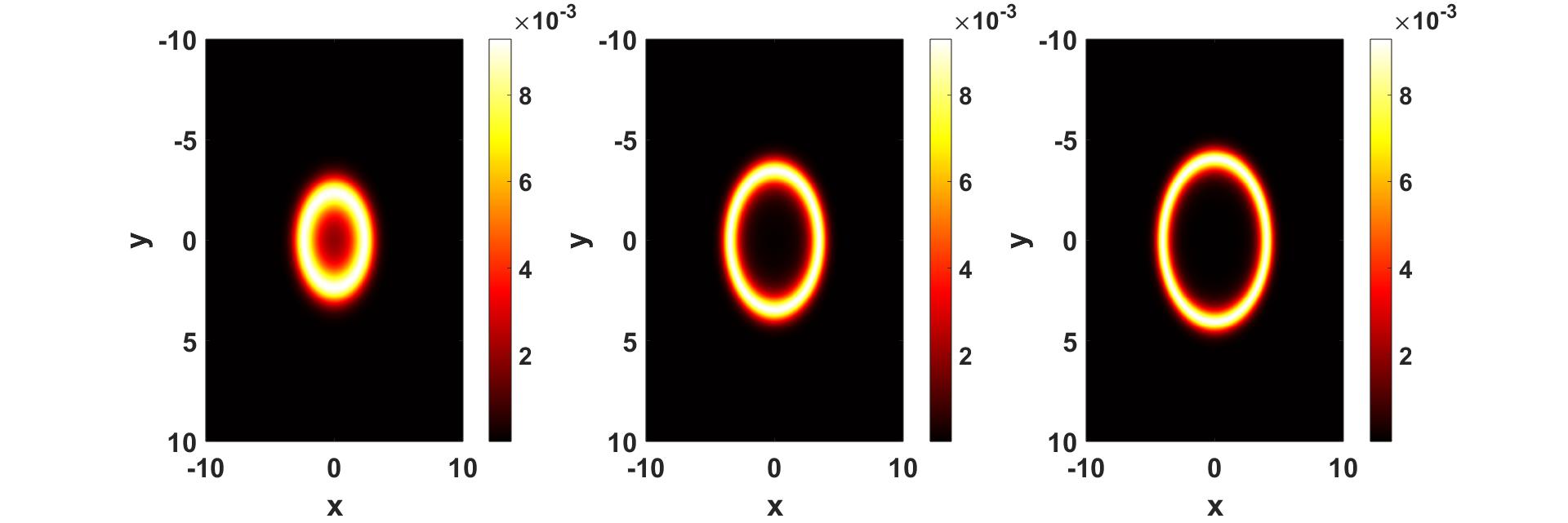}
    \caption{}
    \label{4b}
\end{subfigure}
\caption{a) Variation of the $\eta$ as a function of $r/w$ when $\Omega_{01}=10\gamma$ (blue, circled dots), 50$\gamma$ (red, dash), and 150$\gamma$ (green, solid); (b) intensity profile of the generated vortex beam when $\Omega_{01}==10\gamma$, 50$\gamma$, and 150$\gamma$, respectively, and the other parameters $\Omega_2= 0.5\gamma$, and $\omega$=2 mm. We have chosen $l_3$=0 at $Z=L$.}
\end{figure}

Additionally, in Fig.\ref{4b}, we show the intensity profile of the generated Gaussian and HG beams.
Alternatively, one can generate such HG beams, also by increasing the optical depth (results not shown), as mentioned in Sec. IIA. 

%Now to know the losses, we the $(|\Omega_3|/|\Omega_2|)^2$ as a function of (Z/L). From the plot, we can see that we the losses are minimized as we increase the strength of the control field. To know the values of the strength of the control field for which we can minimize the losses, we plot the $(|\Omega_3|/|\Omega_2|)^2$ with respect to the amplitude of the LG control beam.  

 %\begin{table}[h!]
     %\begin{center}
     %\begin{tabular}{ |p{1cm} | p{10cm} | }
     %\hline
      %($l_1$, $l_2)$ & \hspace{1.8cm}Intensity \hspace{1.5cm} Phase \hspace{1.8cm}     Imag($\rho_{32})$\\ \hline
      %(0,0) &  \includegraphics[width=0.6\textwidth, height=30mm]{v(0,0).jpg}
     % \\ \hline
     % (1,0) &  \includegraphics[width=0.6\textwidth, height=30mm]{v(1,0).jpg}
     % \\ \hline
      %(0,1) &\includegraphics[width=0.6\textwidth, height=30mm]{v(0,1).jpg}
     % \\ \hline
      %(1,1) & \includegraphics[width=0.6\textwidth, height=30mm]{v(1,1).jpg}
     % \\ \hline
      %(2,0) & \includegraphics[width=0.6\textwidth, height=30mm]{v(2,0).jpg}
      %\\ \hline
      %\end{tabular}
      %%\end{center}
     % \end{table}
%\begin{figure}[h]
		%\centering
		%\includegraphics[scale=.5]{intensityrho32.png}
		%\caption{the first coulmn is for the intesity, second is for phase of the generated field of rabi frequency $\Omega_3$, and the third is for imag($\rho_{32}$ ) and $\l_1$ and $l_2$ are topological charge of input fields. } 
%\end{figure}
\section{Microwave generation with \texorpdfstring{$\Lambda$}{\265}-TYPE configuration}
\begin{figure}[h]
		\centering
		\includegraphics[scale=.6]{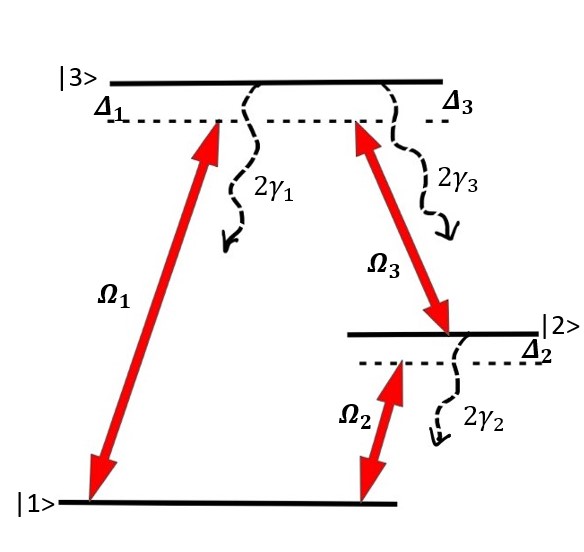}
		\caption{Relevant energy level configuration of the $\Lambda$-type three-level atomic configuration.} 
  \label{fig5}
\end{figure}

Now we select the $\Lambda$-type three-level configuration of an atomic ensemble for the generation of a field in the microwave regime. The relevant energy levels $\ket1$, $\ket2$, and $\ket3$ are shown in Fig. \ref{fig5}. The transitions $\ket1$ $\leftrightarrow$ $\ket3$ and $\ket2$ $\leftrightarrow$ $\ket3$ are coupled via EM field  of frequency $\omega_1$ and $\omega_3$,
%Rabi frequencies $\Omega_1$ and $\Omega_3$,
respectively, in the visible regime. As mentioned before, in the presence of a dc electric field, these types of systems exhibit the absence of inversion symmetry, so that the second-order susceptibility becomes non-zero. This lack of symmetry gives rise to the three-wave mixing and all three transitions becomes dipole-allowed due to the mixing of energy levels of even and odd parity. In this case, we generate a signal of frequency $\omega_2$ = $\omega_1 - \omega_3$ in the microwave range. The phase matching condition is also followed: $\vec{k}_2 = \vec{k}_1 - \vec{k}_3$. 
 $\Omega_i$ is the rabi frequency of the fields, where $i = 1, 2,$ and $3$. 
%If our system is a symmetry-broken quantum system, then don't need to take care of the selection rules of dipole allowed. 

In the context of the $\Lambda$ system, there are two possibilities for choosing the coupling and the control fields. One can either choose using $\Omega_3$ as the control field and $\Omega_1$ as the coupling field or vice versa. In the following, we will comprehensively outline both of these configurations in sequence.

\subsection{\texorpdfstring{$\Omega_3$}{\265} is control field and \texorpdfstring{$\Omega_1$}{\265} is coupling field}
In this case, we consider $\Omega_3\gg \Omega_1$. Since $\Omega_3$ is the stronger field, it will not be affected by the nonlinear interaction.
 To find the coherence $\tilde{\rho}_{21}$ and $\tilde{\rho}_{31}$, we solve Eq. \ref{master} up to the first order of $\Omega_1$, at the steady state. We assume the resonance condition, i.e., $\Delta_1$, $\Delta_2$ and $\Delta_3$ are all zero and also the decay rates $\gamma_{13}$ and $\gamma_{32}$  are equal to $\gamma$, while $\gamma_{12}= 2\gamma$. The relevant coherences thus take the following forms: 
\begin{equation}
\tilde{\rho}_{21} = \frac{-i8\gamma\Omega_2+4\Omega_1\Omega_3^*}{16\gamma^2+5|\Omega_3|^2},\; \tilde{\rho}_{31} = \frac{-i8\gamma\Omega_1+4\Omega_2\Omega_3^*}{16\gamma^2+5|\Omega_3|^2}\;.
\label{rhoL}
\end{equation}
 %Since $\Omega_3$ is the strong field it will not affected by the nonlinear interaction.
 We next solve the following Maxwell's equation only for $\Omega_2$ and $\Omega_1$:
 \begin{equation}
  \frac{\partial\Omega_2}{\partial z} = i\frac{\alpha_{21}\gamma_{21}}{2L}\Tilde{\rho}_{21}^{(1)},  \hspace{.5cm}
  \frac{\partial\Omega_1}{\partial z} = i\frac{\alpha_{31}\gamma_{32} }{2L}\Tilde{\rho}_{31}^{(1)}\;.
  \label{maxl}
\end{equation}
 
 For an initial condition $\Omega_2(z=0) = 0$ and $\Omega_1(z=0)$ = $\Omega(0)$,  the Rabi frequency of the generated field can be obtained as  
 \begin{equation}
 \begin{split}
      \Omega_2(z)  = \frac{-i\Omega_1(0)\Omega_3^*}{\sqrt{\gamma^2-2|\Omega_3|^2}}\Biggl[\exp\left(\frac{2\alpha\gamma\left(3\gamma-\sqrt{\gamma^2-2|\Omega_3|^2}\right)z}{L(16\gamma^2+5|\Omega_3|^2)}\right) 
      \\-\exp\Biggl(\frac{2\alpha\gamma\left(3\gamma+\sqrt{\gamma^2-2|\Omega_3|^2}\right)z}{L(16\gamma^2+5|\Omega_3|^2)}\Biggl)\Biggl].
      \label{solmax}
 \end{split}
 \end{equation}
 \subsubsection{Efficiency}
\begin{figure}
\centering
\begin{subfigure}{1\linewidth}
    \includegraphics[width=7cm]{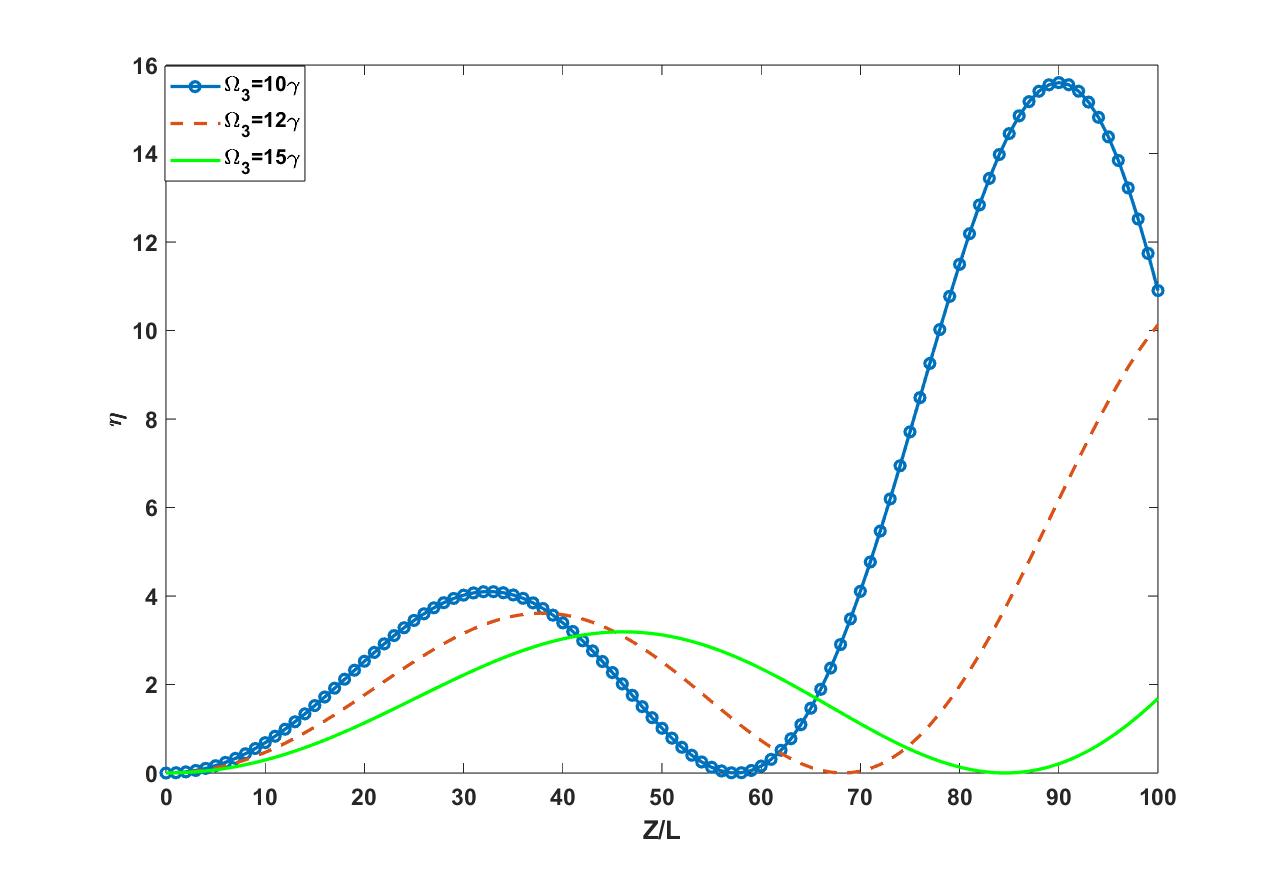}
    \caption{}
    \label{6a}
\end{subfigure}
\hfill
\begin{subfigure}{1\linewidth}
    \includegraphics[width=7cm]{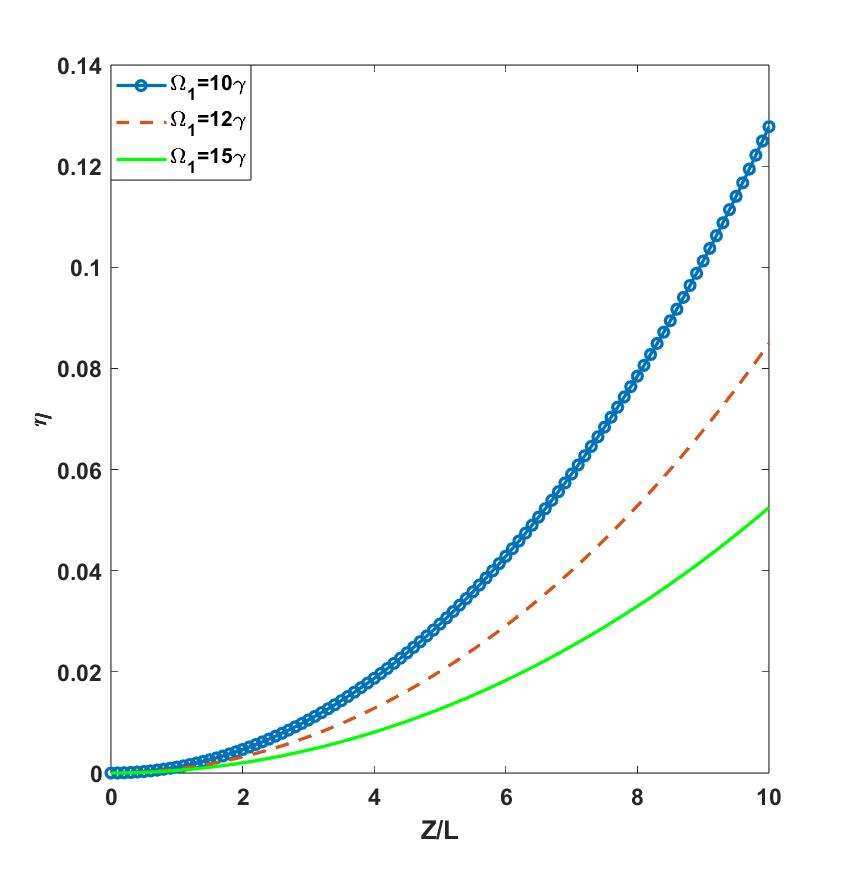}
    \caption{}
    \label{6b}
\end{subfigure}
\caption{Variation of $\eta$ as a function of $Z/L$, when (a) $\Omega_3$ is the control field for $\Omega_3$ = 10$\gamma$ (blue, dots), 12$\gamma$ (red, dash), 15$\gamma$ (green, solid) and (b) $\Omega_1$ is the control field for $\Omega_1$ = 10$\gamma$ (blue, dots), 12$\gamma$ (red, dash), 15$\gamma$ (green, solid).}
\end{figure}

As mentioned before, the generation efficiency is defined as the ratio of the intensities of the generated field and the input field. So, in this case, the efficiency $\eta$ is given by:
 \begin{eqnarray}
     &&\eta =\left(\frac{|\Omega_2(z)|}{|\Omega_1(0)|}\right)^2\nonumber\\
    &&= \frac{|\Omega_3|^2}{\gamma^2-2|\Omega_3|^2}\Biggl[\exp\left(\frac{2\alpha\gamma\left(3\gamma-\sqrt{\gamma^2-2|\Omega_3|^2}\right)z}{L(16\gamma^2+5|\Omega_3|^2)}\right) \nonumber\\
    && - \exp\Biggl(\frac{2\alpha\gamma\left(3\gamma+\sqrt{\gamma^2-2|\Omega_3|^2}\right)z}{L(16\gamma^2+5|\Omega_3|^2)}\Biggl)\Biggl]^2
      \label{etaL}
 \end{eqnarray}
 The term $\sqrt{\gamma^2 - 2|\Omega_3|^2}$ plays a central role in the expression for $\eta$. Given that $\Omega_3$ represents the strong field, we consistently have $\gamma \ll \Omega_3$. Consequently, this term becomes imaginary, leading to oscillatory behavior in $\eta$ (as illustrated in Fig. \ref{6a}). This oscillation extends to the intensity of the generated field within the medium.

The system exhibits an energy exchange between the generated and coupling fields, resulting in oscillations in the intensity of the generated field. By appropriately selecting the value of $Z/L$, we can optimize the intensity to achieve a maximum value.
\subsubsection{Transfer of OAM from coupling to the Generated field}
Form the Eq. (\ref{solmax}) we can see that $\Omega_2$ is proportional to the term $\Omega_1(0)\Omega^*_3$. So the OAM of both fields $\Omega_1$ and $\Omega_3$ can be transferred to the generated field $\Omega_2$. In this section, we consider the coupling field as an LG beam. The phase singularity that is carried by the coupling field is transferred to the generated microwave field, and the field of the topological charge $l_2=l_1$ comes out from the medium. %In table \ref{t2} we show the intensity pattern, phase profile, and imag($\tilde{\rho}_{21}$) of this generated field at Z=L for $l_1$= 1, 2, and 3. From Eq. \ref{solmax}, we can see that \(\Omega_2\) is proportional to the term \(\Omega_1(0)\Omega^*_3\), indicating that the OAM of both fields \(\Omega_1\) and \(\Omega_3\) can be transferred to the generated field. In this section, we consider the coupling field as an LG beam. The phase singularity carried by the coupling field is transferred to the generated microwave field, and the topological charge \(\ell_2 = \ell_1\) emerges from the medium. In Table \ref{t2}, we present the intensity pattern, phase profile, and \(\text{Im}(\tilde{\rho}_{21})\) of this generated field at \(Z = L\) for \(\ell_1 = 1\), 2, and 3. 
In the first column of table \ref{t2}, we display the intensity profile of the generated field, which exhibits a doughnut-like pattern. In the second column, the variation of phase in transverse direction corresponding to the $\ell_1=1, 2,$ and $3$ is displayed. The third column presents \(\text{Im}(\rho_{21})\), where we observe \(2\ell_1\) numbers of petals, revealing the absorption pattern of the generated field in the transverse direction.
%So in the first column of the table \ref{t2} the generated field shows a donut-like intensity pattern according to $\l_1$ value , in the second column we show the variation of phase, and in the third column for imag(\rho_21) we get 2$l_1$ numbers of petals which shows absorption pattern of the generated field in the transverse direction.    
\begin{table*}[ht!]
     \begin{center}
     \begin{tabular}{ |p{.7cm} | p{15cm} | }
     \hline
      ( $l_1)$ & \hspace{3cm}Intensity \hspace{3cm} Phase \hspace{3.5cm}     Im($\rho_{21})$\\ \hline
      1 & \hspace{0cm} \includegraphics[width=0.8\textwidth, height=45mm]{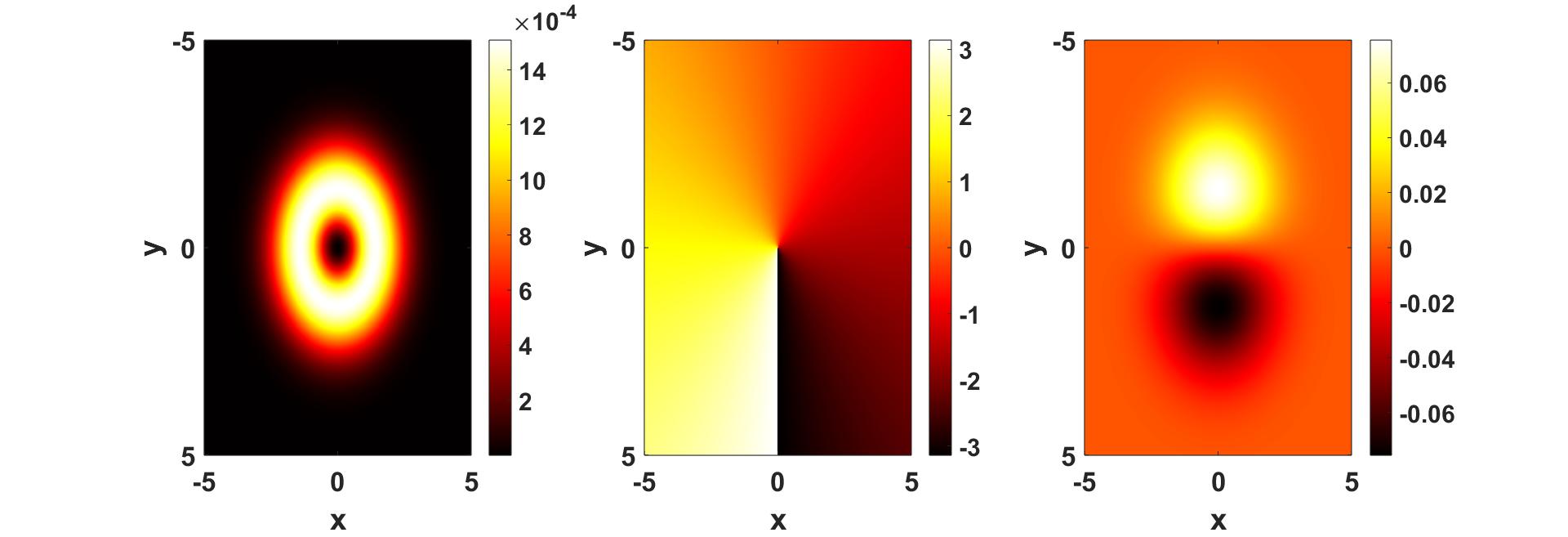}
      \\ \hline
      2 &  \includegraphics[width=0.8\textwidth, height=45mm]{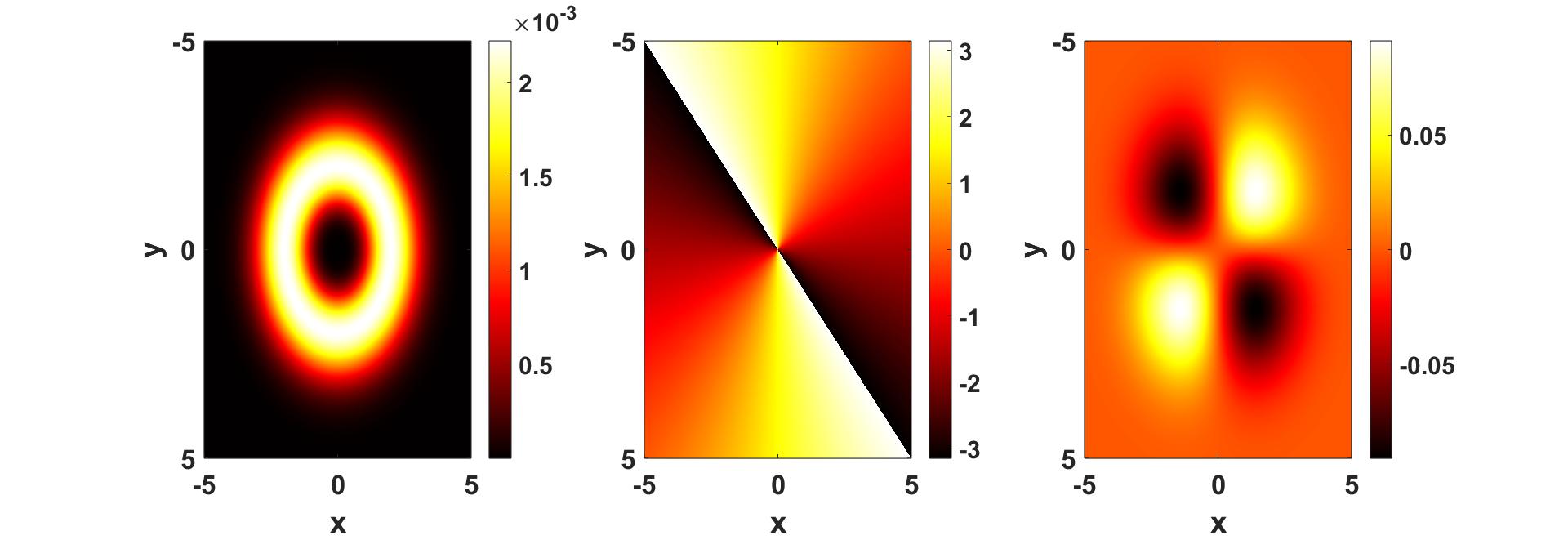}
      \\ \hline
      3 &\includegraphics[width=0.8\textwidth, height=42mm]{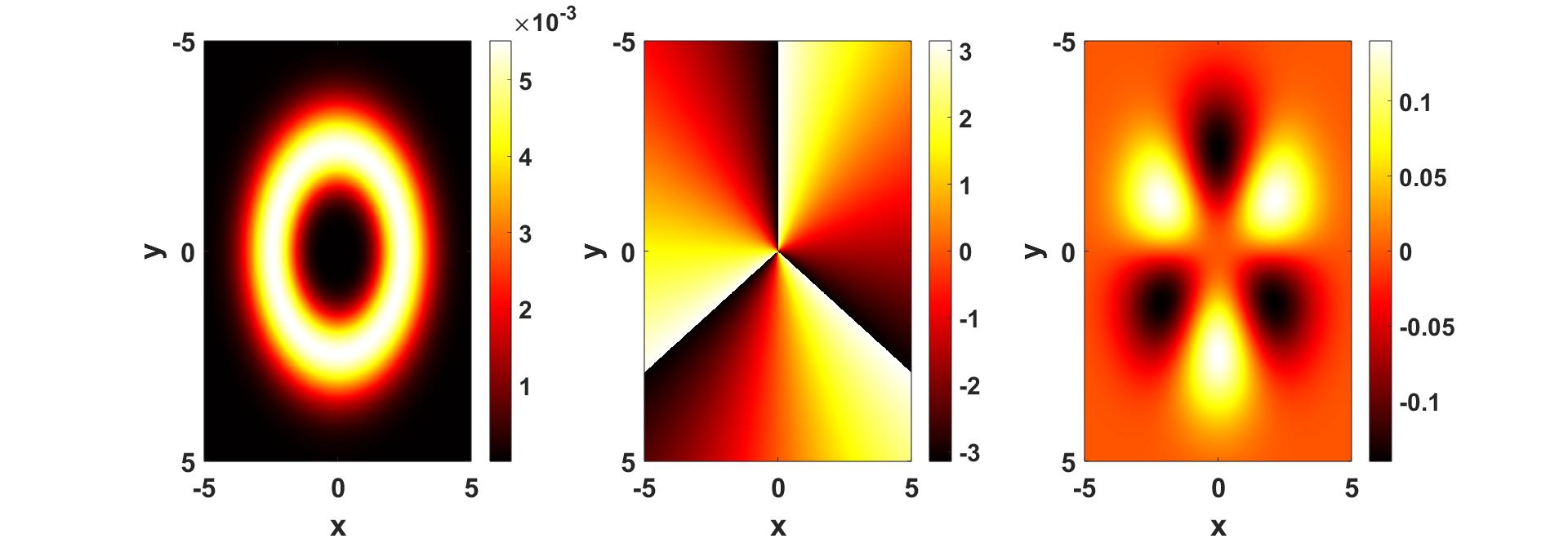}
      \\ \hline
      \end{tabular}
      \caption{Intensity (first column),  phase profile (second column) and, imag($\rho_{21}$) (third column) of the generated field of the Rabi frequency $\Omega_2$ for different modes of coupling LG field, i.e, $l_1$ = 1, 2, and 3. The parameters are $\Omega_3=5\gamma$, $\Omega_{01}=0.5\gamma$, $w= 2$ mm, and $Z=L$.}
      \label{t2}
      \end{center}
      \end{table*}

\subsubsection{Transfer of OAM from control field to Generated field}
\begin{figure*}[ht!]
\centering
\begin{subfigure}{.49\textwidth}
    \includegraphics[width=8cm]{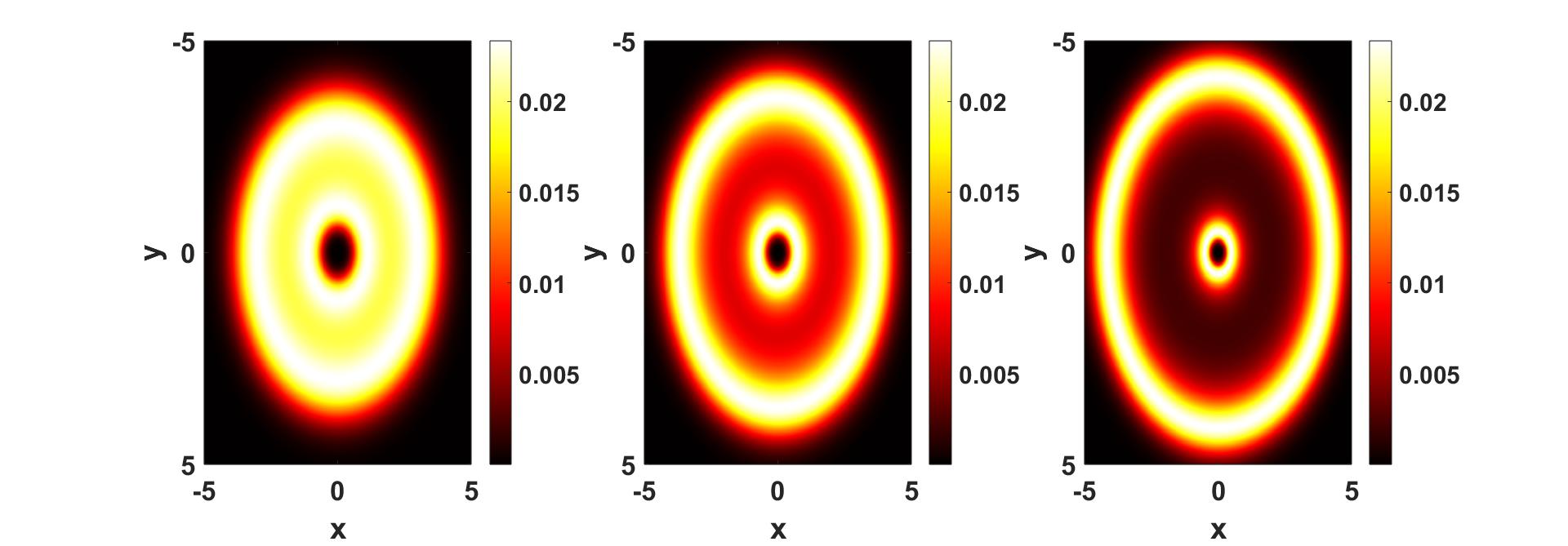}
    \caption{}
    \label{7a}
\end{subfigure}
\hfill
\begin{subfigure}{.49\textwidth}
    \includegraphics[width=8cm]{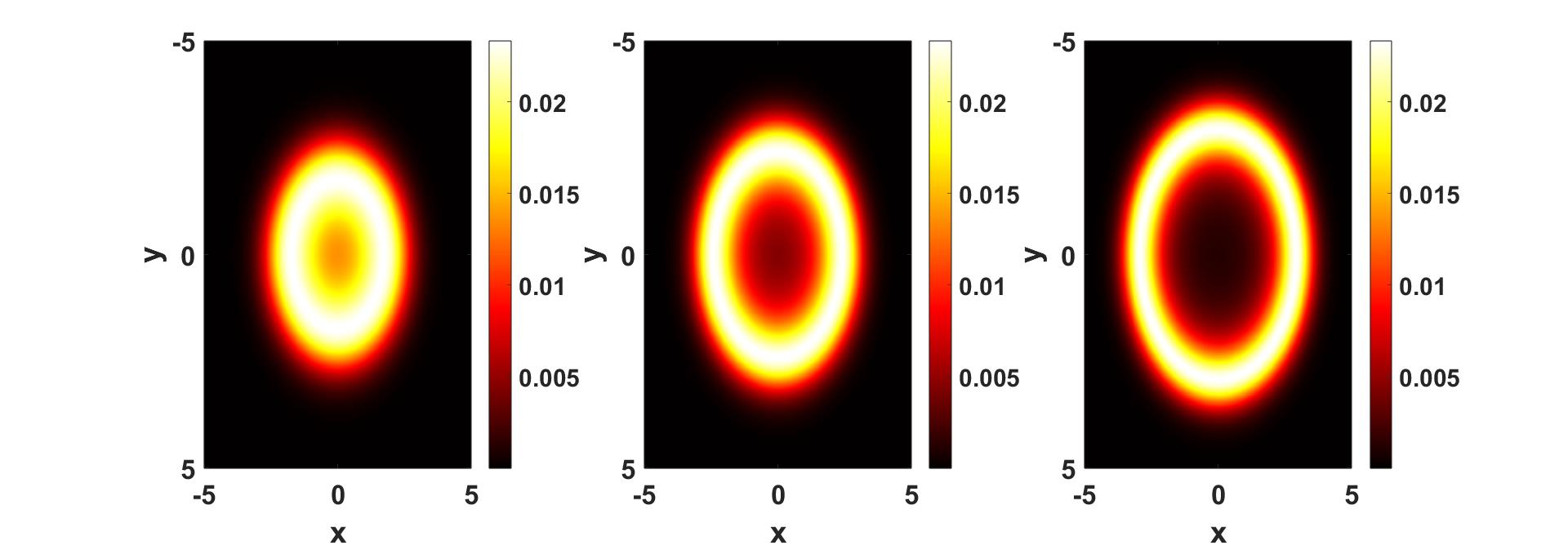}
    \caption{}
    \label{7b}
\end{subfigure}
\caption{Intensity profile of the generated vortex beam for $\Omega_{03}=5\gamma$, 10$\gamma$, 20$\gamma$ respectively, (a) for $l_1$=1 and (b) for $l_1$=0. The parameters are $\Omega_{1}=0.5\gamma$, $w= 2$ mm, and $Z=L$.}
\end{figure*}

Next, we consider the control field to have an LG intensity profile. As $\Omega_2\propto \Omega_1(0)\Omega_3^*$, one can transfer its OAM to the generated beam, which becomes an LG beam with the opposite sign of the topological charge of the control field, while its intensity pattern depends on the strength of the control field. For example, if the intensity of the control field is not strong enough, one can get only one doughnut around an intensity null.  Upon the strength of the control field ($l=1$), one ring of the generated field splits into two (see Fig. \ref{7a}). As in the case of V-type configuration, one can also construct the HGB by applying the Gaussian control field ($l=0$) of high intensity (see Fig. \ref{7b}).

\subsection{ When \texorpdfstring{$\Omega_1$}{\265} is control field and \texorpdfstring{$\Omega_3$}{265} is coupling field}
In this case, $\Omega_1 \gg \Omega_3$. We perturbatively solve Eq. (\ref{master}) in steady state up to the first order of $\Omega_3$ and obtain the following analytical expressions for the coherences $\tilde{\rho}_{21}$ and $\tilde{\rho}_{32}$:
\begin{equation}
  \Tilde{\rho}_{21}^{(1)} = \frac{-4i\gamma\Omega_2+2\Omega_1\Omega^*_3}{8\gamma^2-3|\Omega_1|^2} ,\hspace{.4cm}
   \Tilde{\rho}_{32}^{(1)} = \frac{-2\Omega_1\Omega_2 }{8\gamma^2-3|\Omega_1|^2}\;.
   \label{rhoL2}
\end{equation}
 
For the propagation of the fields $\Omega_2$ and $\Omega_3$ into the medium, we solve the following Maxwell's equations:
 \begin{equation}
  \frac{\partial\Omega_2}{\partial z} = i\frac{\alpha_{21}\gamma_{21}}{2L}\Tilde{\rho}_{21}^{(1)},  \hspace{.5cm}
  \frac{\partial\Omega_3}{\partial z} = i\frac{\alpha_{32}\gamma_{32} }{2L}\Tilde{\rho}_{32}^{(1)}\;.
\end{equation}
with the initial conditions $\Omega_3(z=0)=\Omega_3(0)$ and $\Omega_2(z=0)=0$. We get the following expression for $\Omega_2$:
\begin{equation}
    \begin{split}
        \Omega_2(z) = \frac{i\Omega_3^*(0)\Omega_1}{2\sqrt{\gamma^2+4|\Omega_1|^2}}\Biggl[\exp\Biggl(\frac{\alpha\gamma\left(\gamma-\sqrt{\gamma^2+4|\Omega_1|^2}\right)z}{L(8\gamma^2-3|\Omega_1|^2)}\Biggl) \\
       -\exp\Biggl(\frac{\alpha\gamma\left(\gamma+\sqrt{\gamma^2+4|\Omega_1|^2}\right)z}{L(8\gamma^2-3|\Omega_1|^2)}\Biggl)\Biggl]
      \label{o2L}
     \end{split}
\end{equation}

\subsubsection{Efficiency}
 The above Eq. (\ref{o2L}) indicates that $\Omega_2$ grows as $z$ increases, without any oscillating or damping terms. In contrast, in the first case, when $\Omega_3$ is the control field, $\Omega_2$ yields an oscillatory solution. This amplification effect is illustrated in Fig. \ref{6b}, in which we have displayed the variation of $\eta$ with the normalized propagation distance, alongside the amplification of the coupling field.
%When we introduce a control field between states $\ket1$ and $\ket3$, and a coupling field between states $\ket2$ and $\ket3$, generating a field with a frequency through the difference frequency method,In this case, we observe amplification in the generated field. This phenomenon is absent in the first case, where the control field is applied between states $\ket2$ and $\ket3$. 
Since this configuration satisfies the conditions for parametric amplification, the generated field experiences significant amplification within a medium characterized by a $\Lambda$-type energy level configuration.

\subsubsection{Transfer of OAM to the generated field}
As stated in Eq. (\ref{o2L}), $\Omega_2$ varies in direct proportion to $\Omega_3^*(0)\Omega_1$. This implies that the phase singularity present in either the coupling or the control field can be transferred to the generated field, similar to our approach in the V-system and the first case of the $\Lambda$-type system.

\section{Controlled-NOT gate using OAM states}
 The results obtained for the $\Lambda$ configuration can be interpreted as a controlled-NOT (CNOT) gate in the OAM basis \cite{r30} of three fields. As we have observed, these fields satisfy the OAM conservation $l_{2} = l_{1} - l_{3}$.
 %Lei Wang et al. \cite{r30} demonstrated the construction of a  gate using a double-$\Lambda$ configuration, requiring four fields. In contrast, our model proposes a theoretical AND gate using a V-system or a $\Lambda$-system with only two input fields, under the phase matching and OAM conservation. 
 This means that we have $l_2 = 0$ if $l_1$ and $l_3$ are equal and $l_2$ is 1 if $l_1$ and $l_3$ differ by unity. Considering $l_1$ as the control qubit and $l_2$ as the target qubit, we can identify the following CNOT logic in the basis $|l_1,l_3\rangle \rightarrow |l_1,l_2\rangle$:
 %\begin{equation}
      \begin{eqnarray}
     &&\ket{00}\longrightarrow \ket{00}\nonumber\\
 &&\ket{01}\longrightarrow\ket{01}\nonumber\\
 &&\ket{10}\longrightarrow\ket{11}\nonumber\\
 &&\ket{11}\longrightarrow\ket{10}\nonumber
 \end{eqnarray}
 % \end{equation}
 In Table \ref{t3}, we have shown all of the above four logic operations of the CNOT in the OAM basis using $\Lambda$-configuraion. 
\begin{table*}[ht!]
     \begin{center}
     \begin{tabular}{ |p{1.2cm} | p{15cm} | }
     \hline
      ( S.No.) & \hspace{3cm} $l_1$ \hspace{3.8cm} $l_3$ \hspace{3.5cm}$l_2$
      \\ \hline
      1 & \hspace{0cm} \includegraphics[width=0.8\textwidth, height=40mm]{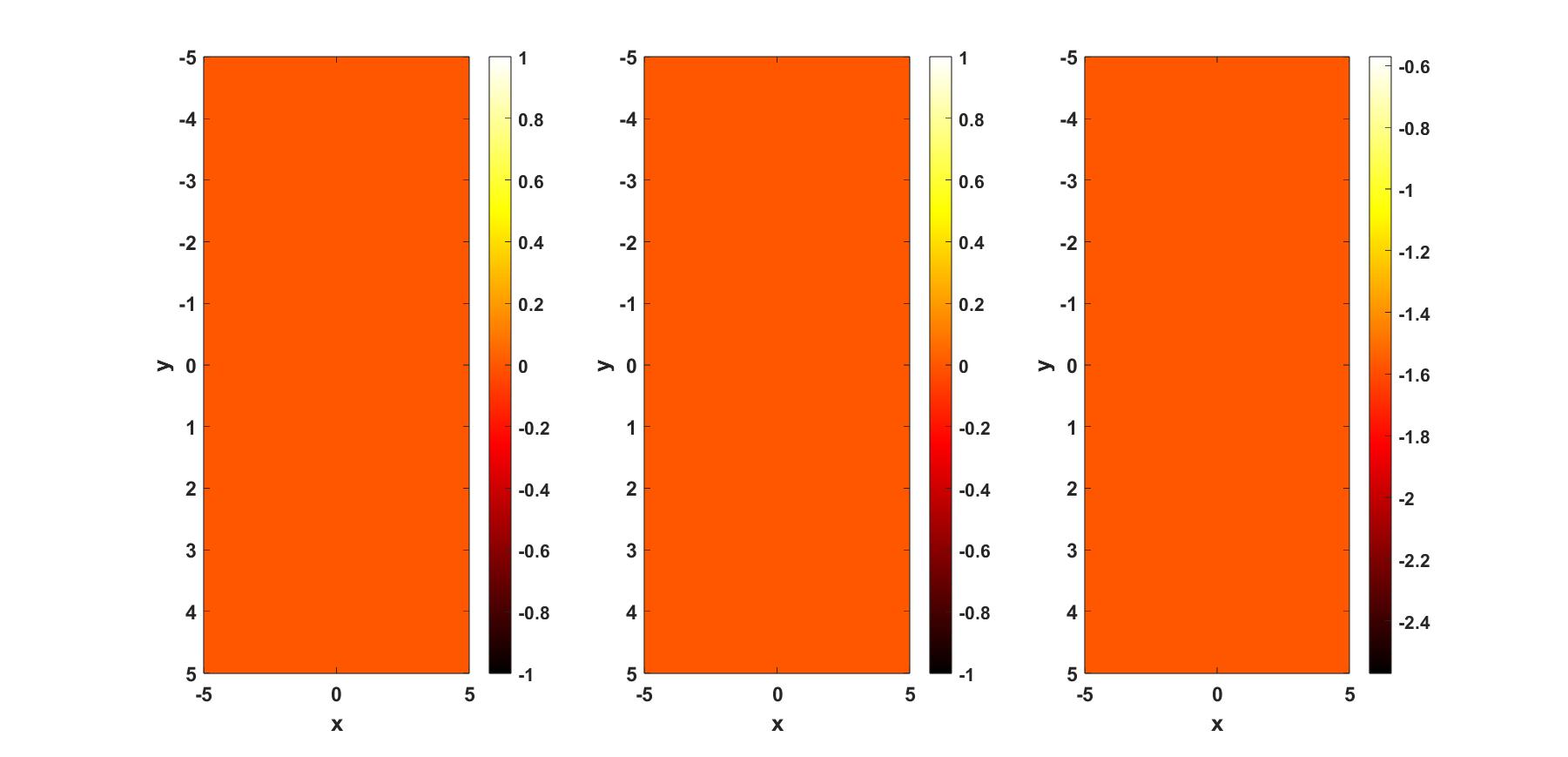}
      \\ \hline
      2 &  \includegraphics[width=0.8\textwidth, height=40mm]{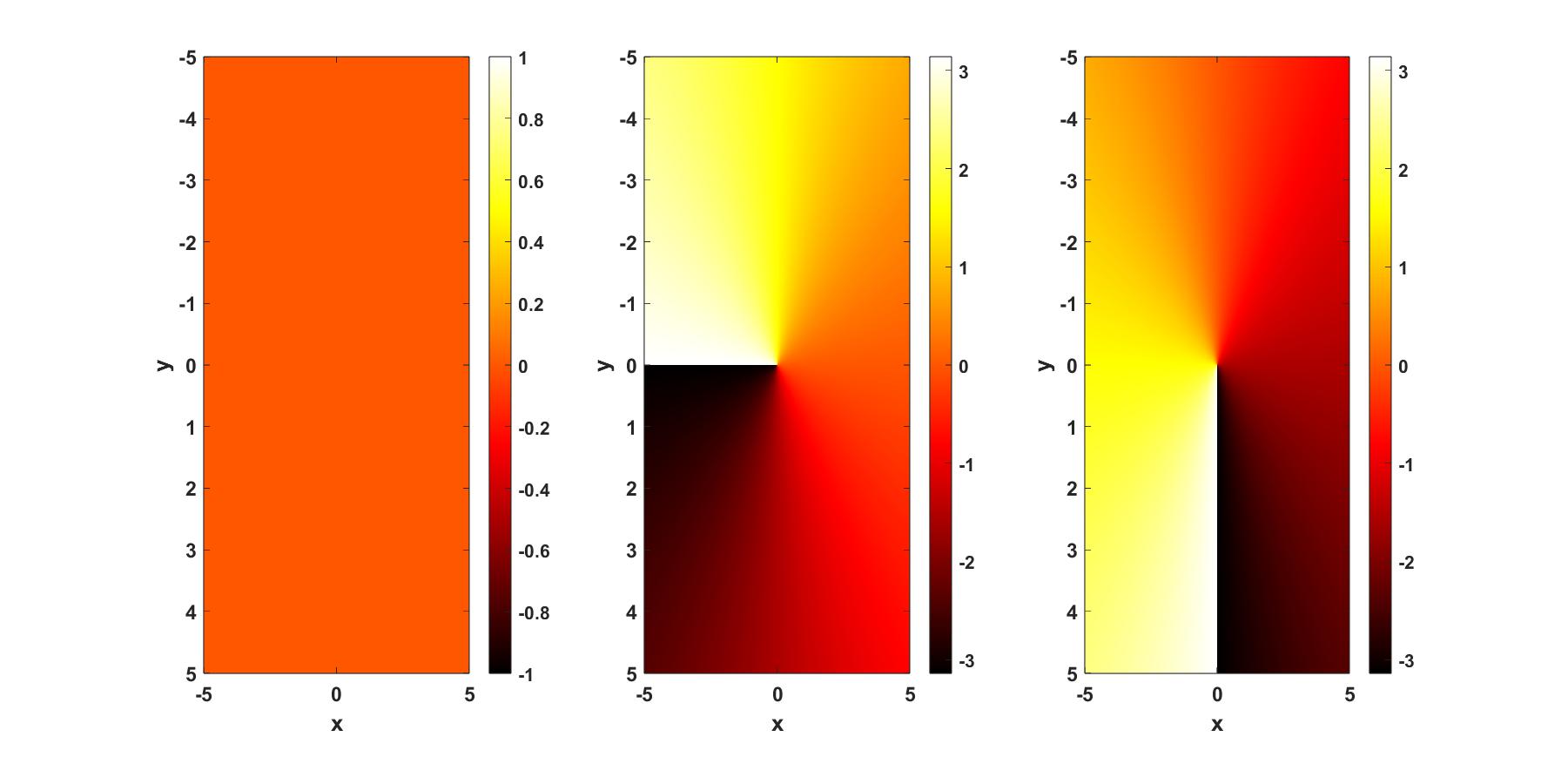}
      \\ \hline
      3 &\includegraphics[width=0.8\textwidth, height=40mm]{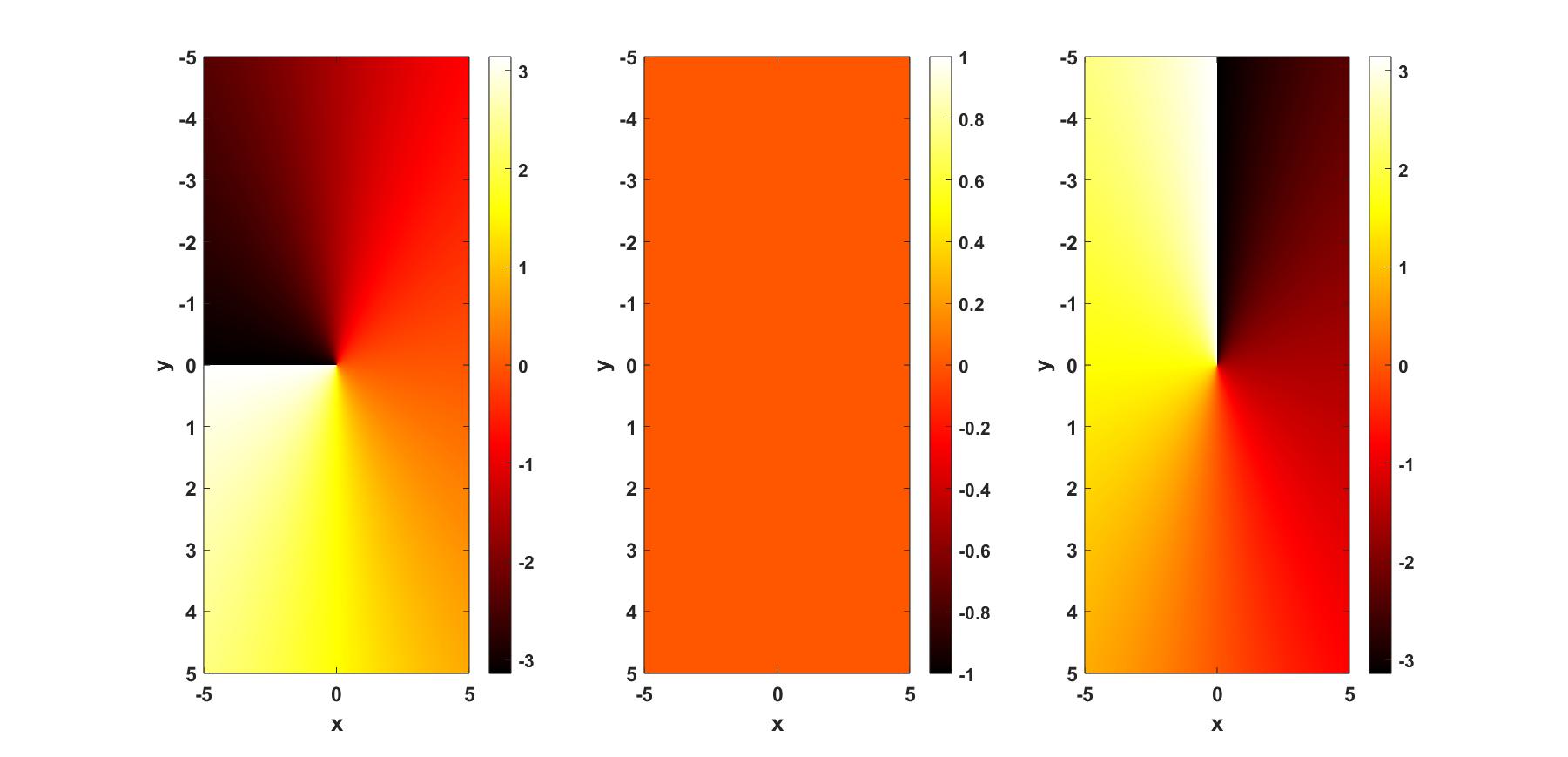}
      \\ \hline
      4 &\includegraphics[width=0.8\textwidth, height=40mm]{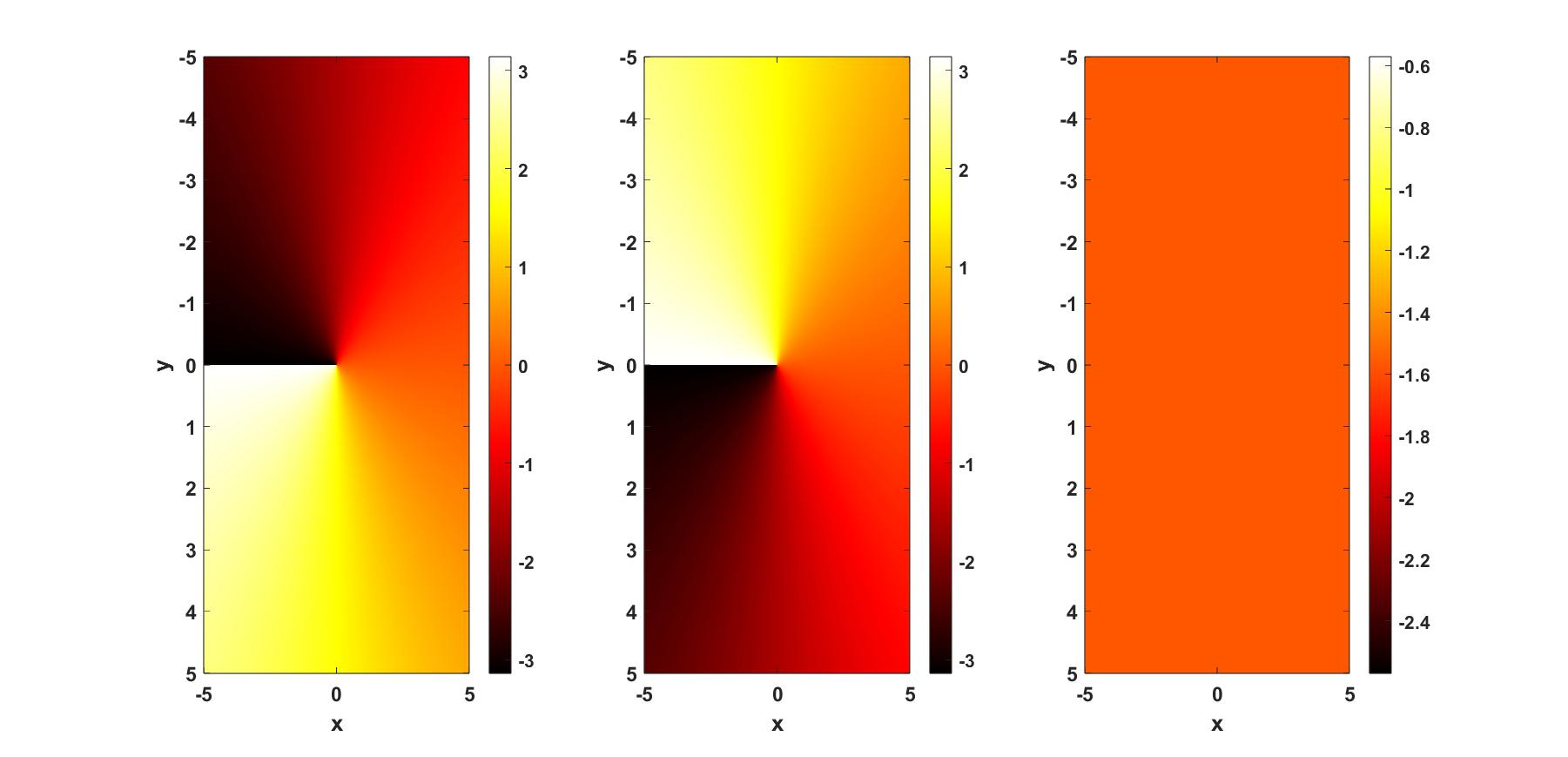}
      \\ \hline
      \end{tabular}
      \caption{CNOT gate using $\Lambda$-system. The ($l_1,l_3$) are (0,0) (first row), (0,1) (second row), (1,0) (third row), and (1,1) (fourth row), with the corresponding outputs $l_2=0, 1, 1, 0$, respectively. The parameters are $\Omega_{03}=0.5\gamma$, $\Omega_{01}=0.5\gamma$, $w= 2$ mm, and $Z=L$.}
      \label{t3}
      \end{center}
      \end{table*}

     Similarly, in the V-configuration, where $l_1$ serves as the control qubit, the  transformation $(l_1,l_2)\rightarrow (l_1,l_3)$ meets the criteria for a CNOT gate.

\section{Discussion}
In this study, we propose a method for generating an em field in the microwave range, using the difference frequency generation process. To make the relevant transition dipole-allowed,  one would need to apply a dc electric field to the atomic vapor, causing the mixing of states with even and odd parity and enabling second-order non-linearity. Here, the fine structure splittings are negligible compared to the Stark splittings whenever the magnitude of the applied electric field is in the range of $10^{3}$-10$^{4}$ V/cm \cite{r29}.

To illustrate our proposal of using the second harmonic generation, we provide two examples involving hydrogenic states. In the first example, the state $\ket{2,1,1}$ (in the $\ket{n,l,m_l}$ basis) can be identified as the ground state, while the states $\frac{1}{\sqrt{3}}(\frac{1}{\sqrt{2}}\ket{3,2,0}-\sqrt{\frac{3}{2}}\ket{3,1,0}+\ket{3,0,0})$ and $\frac{1}{\sqrt{3}}(\frac{1}{\sqrt{2}}\ket{3,2,0}+\sqrt{\frac{3}{2}}\ket{3,1,0}+\ket{3,0,0}$) can be chosen as the excited ones to make the V-configuration. The energy separation between the excited states is $6E_0$, where $E_0$=$3ea_0\epsilon$, with $e$ representing the charge of an electron, $a_0$ denoting the Bohr radius, and $\epsilon$ indicating the strength of the dc electric field. For $\epsilon\sim 10^3$ V/cm,  the transition between these excited states corresponds to the microwave range (with a wavelength $\sim 1.30$ cm), while the other transitions belong to the optical domain. One of the optical fields, carrying OAM, transfers it to the generated microwave em field. 

Our second example corresponds to a $\Lambda$ configuration: the ground states with an energy separation of $2E_0$ are $\frac{1}{\sqrt{2}}(\ket{2,0,0}-\ket{2,1,0)}$ and $\frac{1}{\sqrt{2}}(\ket{2,0,0}+\ket{2,1,0)}$, while the state $\frac{1}{\sqrt{2}} (\ket{3,2,-1}+\ket{3,1,-1)}$ can be chosen as the excited state. In this case, the microwave em field 
 of wavelength 1.47 mm can be generated by the transition between two low-lying states, if the strength of the electric field is $10^3$ V/cm.

 Note that, due to the limiting value of the dc electric field, we are currently able to generate only higher-frequency microwaves. For lower-frequency microwaves, one needs to explore alternative methods.
\section{Conclusions}
We have investigated how to generate a microwave field with an OAM using two different three-level configuration of atomic vapor. We also identify our result with a CNOT gate in the OAM basis. We proposed use of three-wave mixing process, in which two optical fields together create the microwave field. One can introduce a phase singularity in the microwave field, by choosing suitable OAMs in the optical fields.  

In the case of $\Lambda$ configuration, the efficiency of generating the microwave field can be either exponentially increasing or somewhat oscillatory with the propagation length, depending upon the choice of the transition in which the control field is applied.  On the other hand, for a V-configuration, the field gets parametrically amplified as it propagates deeper into the medium.  
%propagation distance demonstrates that there is parametric amplification in the V-system while the $\Lambda$-system efficiency depends on the choice of the control field if $\Omega_3$ is the control field plot shows the oscillatory function of propagation length for the generated field and for the case when $\Omega_1$ is control field efficiency shows exponential growth with respect to the propagation length and we get maximum amplification by selecting the control field of appropriate strength, this can minimize the losses into the medium. \\
Further, by changing the strength of the Gaussian control field, the rings of the LG beam get split, leading to a HGB. 
%This model successfully demonstrates the generation of LG beams and the construction of AND gate in the microwave regime.

To summarize, our proposed method achieves the generation of an LG beam in the microwave regime through a series of steps involving difference frequency generation, phase singularity, and intensity manipulation. Additionally, we can utilize these LG fields of orthogonal OAM states for quantum computing and quantum information processing. %The results demonstrate the feasibility and efficiency of the approach, showcasing its potential for practical applications. 
%Microwave LG fields have the advantage of handling large amounts of information, making microwave LG beams highly valuable for quantum information processing and data storage applications in wireless communication.
Our research also holds promise for future advancements in the field of microwave communication. 
%\section{Appendix}
%\bibliographystyle{plain}

\bibliography{bibfile}

\end{document}